\documentclass[5p,times]{elsarticle}
\usepackage{graphicx} 
\usepackage{xcolor}
\usepackage{acronym}
\usepackage{amsmath}
\usepackage{amssymb}
\usepackage{array}
\usepackage{balance}
\usepackage{pifont}
\usepackage{subcaption}
\usepackage{wasysym} 
\usepackage[hang,flushmargin]{footmisc}
\usepackage{multirow} 
\usepackage{wasysym} 
\usepackage{makecell} 
\usepackage{xr-hyper}
\usepackage{hyperref}

\DeclareMathOperator*{\argmin}{arg\,min}

\acrodef{UAV}{Unmanned Aerial Vehicle}
\acrodef{BER}{Bit-Error-Rate}
\acrodef{PDR}{Packet Delivery Rate}
\acrodef{SNR}{Signal to Noise Ratio}
\acrodef{SDR}{Software-Defined Radio}
\acrodef{RSS}{Received Signal Strength}
\acrodef{PHY}{Physical}
\acrodef{BPSK}{Binary Phase-Shift Keying}
\acrodef{QPSK}{Quadrature Phase-Shift Keying}
\acrodef{QAM}{Quadrature Amplitude Modulation}
\acrodef{OFDM}{Orthogonal Frequency Division Multiplexing}
\acrodef{NN}{Neural Network}
\acrodef{ANN}{Artificial \ac{NN}}
\acrodef{CNN}{Convolutional \acp{NN}}
\acrodef{AI}{Artificial Intelligence}
\acrodef{ML}{Machine Learning}
\acrodef{DL}{Deep Learning}
\acrodef{RJP}{Relative Jamming Power}
\acrodef{ROR}{Receiver Oversampling Ratio}
\acrodef{JOR}{Jamming Oversampling Ratio}
\acrodef{GCS}{Ground Control Station}
\acrodef{RPAS}{Remotely-Piloted Aircraft System}
\acrodef{AWGN}{Additive White Gaussian Noise}
\acrodef{AUC}{Area Under the Curve}
\acrodef{ROC}{Receiver Operating Characteristic}
\acrodef{TPR}{True Positive Ratio}
\acrodef{TNR}{True Negative Ratio}
\acrodef{AGC}{Automatic Gain Control}
\acrodef{WSN}{Wireless Sensor Networks}
\acrodef{IoT}{Internet of Things}
\acrodef{CPU}{Central Processing Unit}
\acrodef{PLR}{Packet Loss Ratio}
\acrodef{PRR}{Packet Rejection Ratio}
\acrodef{PDR}{Packet Delivery Ratio}
\acrodef{CER}{Chip Error Rate}
\acrodef{STFT}{Short-Time Fourier Transform}
\acrodef{CGDWT}{Complex Gaussian Derivative Wavelet Transform}
\acrodef{ML}{Machine Learning}
\acrodef{CNN}{Convolutional Neural Network}
\acrodef{C-GAN}{Conditional Generative Adversarial Network}
\acrodef{DBN}{Dynamic Bayesian Network}
\acrodef{HMM}{Hidden Markov Model}
\acrodef{RF}{Radio Frequency}
\acrodef{K-NN}{K-Nearest Neighbors}
\acrodef{HHT}{Heuristic Hypothesis Test}
\acrodef{DCNN}{Deep Convolutional Neural Network}
\acrodef{ST}{Stockwell Transform}
\acrodef{BIM}{Building Information Modeling}
\acrodef{MSE}{Mean Squared Error}
\acrodef{IQ}{In-Phase - Quadrature}
\acrodef{HPC}{High-Performance Computing}
\acrodef{AE}{Autoencoder}
\acrodef{COTS}{Commercial Off-The-Shelf}
\acrodef{sps}{samples per second}

\newcolumntype{P}[1]{>{\centering\arraybackslash}p{#1}}

\journal{Computer Networks}

\begin{document}
\begin{frontmatter}

\title{Weak-Jamming Detection in IEEE 802.11 Networks: \\ Techniques, Scenarios and Mobility}


\tnotetext[t1]{This work has been partially supported by the INTERSECT project, Grant No. NWA.1162.18.301, funded by the Netherlands Organization for Scientific Research (NWO). Any opinions, findings, conclusions, or recommendations expressed in this work are those of the author(s) and do not necessarily reflect the views of NWO. Moreover, this publication was made possible by the NPRP12C-0814-190012-SP165 award from the Qatar National Research Fund (a member of Qatar Foundation).}
\author[1]{Martijn Hanegraaf%
}
\ead{m.f.g.hanegraaf@student.tue.nl}
\author[1]{Savio Sciancalepore\corref{cor1}}
\ead{s.sciancalepore@tue.nl}
\author[2]{Gabriele Oligeri}
\ead{goligeri@hbku.edu.qa}
\cortext[cor1]{Corresponding author}

\affiliation[1]{organization={Eindhoven University of Technology (TU/e)},
city={Eindhoven},
country={Netherlands}}

\affiliation[2]{organization={Hamad Bin Khalifa University (HBKU), College of Science and Engineering (CSE)},
city={Doha},
country={Qatar}}

\begin{abstract}
    \textcolor{black}{
    State-of-the-art solutions detect jamming attacks \emph{ex-post}, i.e., only when jamming has already disrupted the wireless communication link. In many scenarios, e.g., mobile networks or static deployments distributed over a large geographical area, it is often desired to detect jamming at the early stage, when it affects the communication link enough to be detected but not sufficiently to disrupt it (detection of \emph{weak} jamming signals). Under such assumptions, devices can enhance situational awareness and promptly apply mitigation, e.g., moving away from the jammed area in mobile scenarios or changing communication frequency in static deployments, before jamming fully disrupts the communication link. Although some contributions recently demonstrated the feasibility of detecting low-power and weak jamming signals, they make simplistic assumptions far from real-world deployments. Given the current state of the art, no evidence exists that detection of weak jamming can be considered with real-world communication technologies.\\
    In this paper, we provide and comprehensively analyze new general-purpose strategies for detecting weak jamming signals, compatible by design with one of the most relevant communication technologies used by commercial-off-the-shelf devices, i.e., IEEE 802.11. We describe two operational modes: (i) binary classification via Convolutional Neural Networks and (ii) one-class classification via Sparse Autoencoders. We evaluate and compare the proposed approaches with the current state-of-the-art using data collected through an extensive real-world experimental campaign in three relevant environments. At the same time, we made the dataset available to the public. Our results demonstrate that detecting weak jamming signals is feasible in all considered real-world environments, and we provide an in-depth analysis that considers different techniques, scenarios, and mobility patterns. 
    }
\end{abstract}


\end{frontmatter}


\section{Introduction}
\label{sec:intro} 
Nowadays, jamming is one of the most simple and effective ways to disrupt the operations of wireless networks. In fact, by injecting high-power noise into the communication channel used by \ac{RF} devices, malicious parties disrupt the communications of static and mobile devices, affecting significantly large areas. As a result, devices lose the ability to communicate, leading to networking issues~\cite{pirayesh2022_comst}.

Several solutions are available today for detecting jamming in wireless networks~\cite{wang2020_commag}. However, most of such solutions detect jamming \emph{ex-post}, i.e., after jamming has already affected the quality of the wireless communication link. \textcolor{black}{For instance, consider the scenario where a mobile device (vehicle, robot, or drone) approaches an area under jamming, thus experiencing an increasing effect of the jammer, with a growing \ac{BER}, which, in turn, affects the quality of the communication link. In such a scenario, it is desirable to detect jamming \emph{earlier}, when the jamming signal is weak, and thus, not enough to break the communication link, but strong enough to be detected.
Detection of weak jamming signals is also a valuable technique for large static networks, such as the Internet of Things (IoT), where nodes located on the boundaries of the jammed area can detect the presence of the (weak) jamming signal while not being affected by it---they are still able to communicate with their neighbors while being able to detect the presence of the weak jamming signal. Jamming detection in these scenarios enhances situational awareness, allowing the system administrator to switch to a safer and more robust network configuration or communication protocol. Such scenarios are sometimes referred to as \emph{low-BER regimes} and \emph{low-power} jamming, and to the best of our knowledge, jamming detection solutions suitable for such scenarios and validated with real-world data have been considered only by our preliminary work in~\cite{alhazbi2023_ccnc} and \cite{sciancalepore2024_iotj}. }

Although the results reported in these works demonstrate the feasibility of detecting weak jamming in indoor scenarios, many gaps remain unresolved. In particular, it is unclear whether detection of weak jamming signals could also work effectively and reliably outdoors, where the multipath effect may affect the signal behavior differently than indoors. Moreover, the experiments in~\cite{alhazbi2023_ccnc} and~\cite{sciancalepore2024_iotj} consider only the most straightforward digital modulation technique, i.e., \ac{BPSK}, and a \emph{clean} communication channel. It is unclear how to generalize the solution conceived therein for more complex modulation schemes used by mobile devices in the real world, e.g., IEEE 802.11 (WiFi), working in the crowded $2.4$~GHz frequency band. Finally, the technique designed in~\cite{sciancalepore2024_iotj} requires the acquisition of a significant amount of data for training a \ac{DL} model; no investigation has been conducted so far about the size of the training set and how this affects the performance of the classifier, as well as the effectiveness of data augmentation techniques.  

{\bf Contribution.} In this paper, we improve and extend our previous contributions in~\cite{alhazbi2023_ccnc} and~\cite{sciancalepore2024_iotj} through the design of general-purpose solutions to detect weak jamming signals, compatible by design with real-world digital modulation schemes used by modern \ac{COTS} devices. Our proposed solution converts raw \ac{PHY}-layer signals acquired from the wireless channel into images and then uses various \ac{DL}-based solutions to detect jamming by correct classification of such images. We provide two operational modes of our approach, i.e., binary classification via \acp{CNN} and anomaly detection via \acp{AE}. We evaluated the effectiveness of these approaches for weak-jamming detection through an extensive real-world experimental campaign encompassing three relevant environments (indoor, outdoor with multipath, outdoor with minimal multipath) and several protocol configurations, jamming parameters, and communication parameters. As relevant results, we demonstrate that our newly proposed image generation technique is successful for weak jamming detection, and it generalizes by design with more complex modulation schemes. We also demonstrate that when detecting jamming becomes particularly challenging, \acp{CNN} outperform \acp{AE} thanks to the availability of anomalous (jamming) samples at training time. We also release the dataset collected for this study open-source at~\cite{data}, fostering future research.

We highlight that this work extends further our previous studies in~\cite{alhazbi2023_ccnc} and~\cite{sciancalepore2024_iotj}, through the following new contributions:
\begin{itemize}
    \item We extend further the image generation methodology introduced in the previous papers~\cite{alhazbi2023_ccnc} and~\cite{sciancalepore2024_iotj} through the design of an enhanced image generation technique taking into account the higher complexity of the real-world modulation schemes, and we demonstrate experimentally the advantages and trade-offs connected with such a choice.
    \item Although previous studies evaluated weak-jamming detection only indoors, in this study, we extend our experimental analysis by carrying out extensive data collection outdoors, considering environments characterized by heterogeneous multipath conditions, and we release our data open source.
    \item While previous studies only consider data delivered on custom communication frequencies (e.g., $900$~MHz) through the \ac{BPSK} modulation, in this study, we extend our analysis considering an actual communication technology used by wireless devices outdoors, i.e., IEEE 802.11, working on the crowded $2.4$~GHz frequency band, using the \ac{OFDM} modulation scheme with \ac{BPSK}, \ac{QPSK}, 16-\ac{QAM}, and 64-\ac{QAM}.
    \item We extensively compare the performance of the proposed jamming detection techniques with the approaches previously proposed in~\cite{alhazbi2023_ccnc} and~\cite{sciancalepore2024_iotj} across several configurations, jamming techniques, and communication parameters. 
    \item We investigated the problem of data set size and evaluated the performance of the classifier with different training set sizes using various data augmentation strategies traditionally used for image processing, including rotation, mirroring, contrast adjustment, and brightness adjustment. We show that some strategies contribute to keeping jamming detection accuracy high while requiring fewer real-world data.
\end{itemize}

This research demonstrates experimentally the potential and feasibility of detecting weak jamming signals in many different scenarios and operational conditions, ultimately paving the way for adopting such techniques in the wild.

{\bf Roadmap.} The paper is organized as follows. Sec.~\ref{sec:background} introduces the preliminaries, Sec.~\ref{sec:related} reviews related work, Sec.~\ref{sec:scenario_adv_model} introduces our scenarios and adversarial model, Sec.~\ref{sec:methodology} provides the details of the investigated methodologies for weak-jamming detection, Sec.~\ref{sec:data_experiments} describes our data collection campaign, Sec.~\ref{sec:results} presents the results of our analysis, Sec.~\ref{sec:discussion} discusses relevant results and deployment aspects of our solutions and, finally, Sec.~\ref{sec:conclusion} concludes the paper.
\textcolor{black}{
\section{Preliminaries}
\label{sec:background}
In this section, we introduce preliminary concepts used in our paper, i.e., digital modulation techniques (Sec.~\ref{sec:modulation}) and \ac{DL} strategies for jamming detection (Sec.~\ref{sec:dl}).
\subsection{Digital Modulation}
\label{sec:modulation}
Digital modulation schemes used for modern wireless communications transform a digital bit stream to a \ac{RF} signal, suitable to be transmitted over a wireless channel~\cite{rappaport2024_book}. Such \ac{RF} signal is typically expressed as a complex number of the form $\mathbf{s} = \mathbf{I}+j\mathbf{Q}$, generally denoted as \ac{IQ} samples, where I is the real component and Q is the imaginary component. 
As an example, Fig.~\ref{fig:BG_Modulations} shows (real) IQ constellation plots for the modulation schemes \ac{BPSK}, \ac{QPSK}, 16-\ac{QAM} and 64-\ac{QAM}, respectively, as obtained during the experiments carried out for this work.
\begin{figure}
    \centering
    \includegraphics[width=1.0\columnwidth]{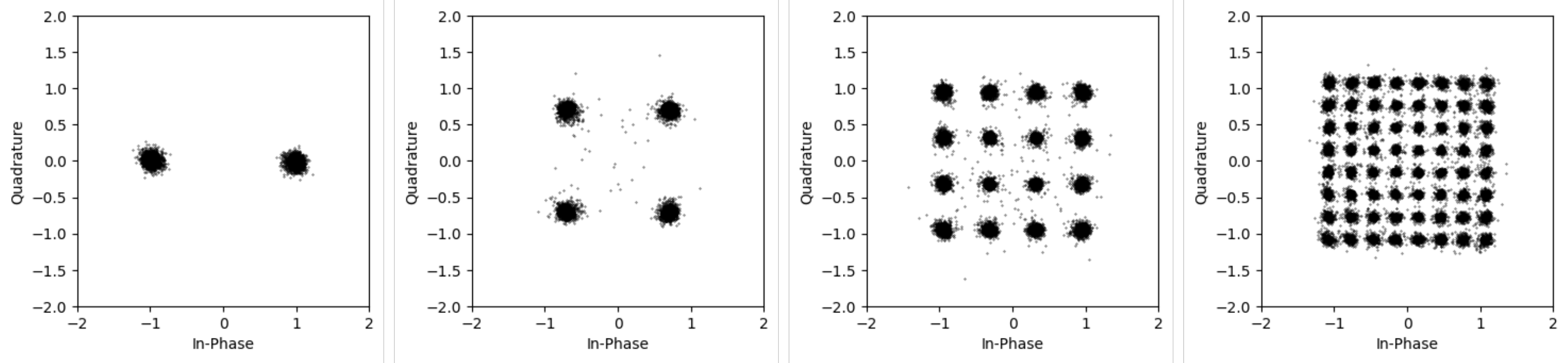}
    \caption{IQ constellations for BPSK, QPSK, 16-QAM and 64-QAM, respectively, obtained from our experiments.}
    \label{fig:BG_Modulations}
\end{figure}
We notice that the higher the order of the modulation, i.e., the number of different \ac{IQ} values, the higher the number of bits that an IQ sample can carry. For example, each IQ sample in the \ac{BPSK} scheme carries a single bit (0 or 1), while an IQ sample in the \ac{QPSK} scheme carries two bits (00, 01, 10, or 11). At the same time, we also notice that the transmission of the IQ samples on the wireless channel affects their value, thus changing the original value of the sample, i.e., moving the position of the sample from the expected location in the IQ constellation plane. 
When the noise is low, and the shift from the expected position is minimal, errors do not affect the mapping. However, the higher the order of the modulation scheme, the greater the chance that the wireless channel moves a sample far away from the expected one, thus causing a \emph{decoding error}. 
Digital modulation schemes trade off performance (throughput) with robustness to noise. 
In this paper, in line with our previous works in~\cite{alhazbi2023_ccnc} and~\cite{sciancalepore2024_iotj}, we consider \emph{Low-\ac{BER} Regime} as a communication channel in which noise slightly affects the quality of the communication link, causing only a few decoding errors, below $1\%$. The techniques described in this paper aim to detect the presence of weak jamming signals in such challenging conditions, i.e., before jamming takes down the communication link.
\subsection{Deep Learning Tools for Jamming Detection}
\label{sec:dl}
We consider two \ac{DL}-based tools for image classification, i.e., \aclp{CNN} and \aclp{AE}.\\
\noindent
{\bf \aclp{CNN}.} \acp{CNN} are primarily used in the field of pattern recognition within images, with the task of classifying them~\cite{guo2016_neuro}. The input of the classifier, usually in the form of a multidimensional vector, is loaded into the input layer, which distributes it over the hidden layers. Without loss of generality, a \ac{CNN} comprises three neural layers, i.e., convolutional layers, pooling layers, and fully connected layers. The convolutional layer generates various feature maps from the input using the mathematical convolution operation, producing one function from two input functions. The feature map is an abstraction of the input image. The pooling layer follows the convolutional layer and reduces the dimensions of the feature maps and network parameters. 
In the fully connected layer, each neuron in the previous layer is connected to each neuron in the current layer. The number of neurons in this layer usually matches the number of output classes \cite{guo2016_neuro}. \acp{CNN} take as input raw image pixels and outputs class scores, indicating the likelihood that the input sample is part of the specific class. Note that training a \ac{CNN} requires images from all classes. \\
\noindent
{\bf Autoencoders.} \acp{AE} are neural networks designed to encode the input into a compressed representation, i.e., the latent representation, and then decode it back so that it is as similar as possible to the input~\cite{bank2021_arxiv}. The problem can be summarized by identifying an encoder function $A : \mathbb{R}^n \rightarrow \mathbb{R}^p$ and a decoder function $B : \mathbb{R}^p \rightarrow \mathbb{R}^n$ that satisfy Eq.~\ref{eq:Autoencoder}:
\begin{equation}
    \label{eq:Autoencoder}
        \argmin_{A,B}E[\Delta(\textbf{x}, B \circ A(\textbf{x}))],
\end{equation}
where $E$ represents the expectation over the distribution of $x$, $\circ$ is the composition operator, and $\Delta$ is the function that measures the distance between the input and the output of the decoder, called the reconstruction loss function.
In the cybersecurity research domain, \acp{AE} have been used successfully for anomaly detection~\cite{feng2022_iotj},~\cite{cook2020_iotj}. Let $J$ and $K$ be two probability distributions such that $J \neq K$ and let $c$ be an \ac{AE} model trained on samples from $J$. When using $c$ to reconstruct samples, we expect a smaller reconstruction error on samples from $J$ than unseen samples from $K$~\cite{oligeri2022_tifs}. We can identify a specific reconstruction error value and use it as a threshold value $\tau$, defining a boundary for classification. Any input sample to $c$ with a reconstruction error greater than $\tau$ can be classified as $\notin J$, i.e., an anomaly. Note that, differently from \acp{CNN}, \acp{AE} require only one class for training, i.e., the samples of the \emph{legitimate} class (no anomalies). In Sec.~\ref{sec:methodology}, we discuss in more detail what such a feature entails for jamming detection.
}

\section{Related Work}
\label{sec:related}

Several jamming detection approaches have been proposed in the scientific literature, as discussed in recent surveys such as~\cite{pirayesh2022_comst} and summarized in Tab.~\ref{tab:related}. 
\begin{table*}[!t]
\caption{Qualitative comparison of relevant literature on jamming detection. Duplicated rows with the same reference indicate multiple proposals by the same paper. For evaluation setup, \CIRCLE\ indicates Indoor and Outdoor evaluation, \LEFTcircle\ indicates Indoor or Outdoor, and \Circle\ indicates simulation or analytical evaluation. The symbol \emph{-} indicates that the paper provides no information about a specific feature. \textcolor{black}{The works~\cite{oligeri2024_sac},~\cite{alhazbi2023_acsac}, and~\cite{papangelo2023commag} are not reported since, although applying image generation from RF signals, they do not deal with jamming detection.}
}
\scriptsize
\color{black}
\begin{tabular}{ c | c | c | c | c | c | c | c | c | c | c}
 \centering
 \textbf{Ref.} & \textbf{Metric} & \textbf{Technique} & \makecell{\textbf{Communication} \\ \textbf{Technology}} & \makecell{\textbf{Modulation} \\ \textbf{Scheme}} & \makecell{\textbf{Jamming} \\ \textbf{Signal}} & \makecell{\textbf{Robustness}\\\textbf{to Distance}} & \makecell{\textbf{Weak}\\\textbf{Jamming}} & \makecell{\textbf{Jamming}\\\textbf{Knowledge}} & \makecell{\textbf{Eval.} \\ \textbf{Setup}} & \makecell{\textbf{Robustness} \\ \textbf{to Mobility}}\\
 \hline
 
 \cite{punal2014_wowmom}      & IT, PDR, RSSI   & RF & IEEE 802.11 & OFDM & BPSK & \Circle & \Circle & \Circle & \LEFTcircle & \CIRCLE \\[0.2cm]
 
 \cite{li2022_access} & Spectrogram     & CNN & IEEE 802.11 & OFDM & \makecell{AWGN, Tone, \\ Pulse, OFDM} & \Circle & \Circle & \Circle & \LEFTcircle & \Circle\\[0.2cm]
 
 \cite{li2022_access} & \makecell{SNR, RSSI, \\ OFDM-Specific} & RF & IEEE 802.11 & OFDM & \makecell{AWGN, Tone, \\ Pulse, OFDM} & \Circle & \Circle & \Circle & \LEFTcircle & \Circle\\[0.2cm]

 \cite{toma2020_tccn} & ST & C-GAN & 5G & \makecell{B/QPSK,\\ 16/64-QAM} & AWGN & \Circle & \Circle & \CIRCLE & \LEFTcircle  & \Circle\\[0.2cm]
 
 \cite{toma2020_tccn} & I/Q & DBN & 5G & OFDM & OFDM & \Circle & \Circle & \Circle & \Circle & \Circle\\[0.2cm]

 \cite{furqan2020_jwcn}  & \makecell{Res. Energy\\ Decay Rate} & ML & - & \makecell{QAM, PAM,\\ FSK, PSK} & AWGN & \Circle & \Circle & \Circle & \Circle & \Circle \\[0.2cm]
 
 \cite{zhang2023_tccn} & \makecell{Log. Received\\ Energy} & HMM & - & - & AWGN & \Circle & \Circle & \CIRCLE & \Circle & \Circle\\[0.2cm]
 
 \cite{arjoune2020_icoin}           & \makecell{BPR, PDR, \\ RSSI, CCA} & RF & 5G & - & AWGN & \Circle & \Circle & \Circle & \LEFTcircle & \Circle\\[0.2cm]

 \cite{spuhler2014_twc}           & CER & HHT & IEEE 802.15.4 & DSSS & DSSS & \Circle & \Circle & \CIRCLE & \LEFTcircle & \CIRCLE\\[0.2cm]

\cite{gecgel2019_wiseml}           & STFT, CGDWT & \makecell{DCNN,\\ DRNN} & - & OFDM & \makecell{AWGN, \\ OFDM} & \Circle & \Circle & \Circle & \LEFTcircle & \Circle\\[0.2cm]

\cite{krayani2020_pimrc}           & \makecell{Timing\\Resource Alloc. }& DBN & 4G & OFDM & OFDM & \Circle & \Circle & \Circle & \Circle &  \Circle\\[0.2cm]

\cite{kihei2021_wfiot}           & \makecell{Heuristic\\Proc. Features }& K-NN & IEEE 802.11p & OFDM & AWGN & \Circle & \Circle & \Circle & \LEFTcircle & \Circle\\[0.2cm]

\cite{bogdanoski2014_conf}  & RSS & \makecell{Effective\\Radiated\\Power}  & WiMAX & DSSS & \makecell{Pulse\\Sweep\\Single-band} & \Circle & \CIRCLE & \Circle & \Circle & \Circle \\[0.2cm]

\cite{garnaev2016_tsp} & Time & \makecell{Game\\ Theory} & - & -  & AWGN & \Circle & \CIRCLE & \CIRCLE  & \Circle & \Circle \\[0.2cm]

\cite{venkata2021_wlett} & \makecell{Packet\\ Drop\\Probability} & \makecell{SVM,\\ Random Forests} & IEEE 802.11 & OFDM & AWGN & \Circle & \CIRCLE & \Circle & \Circle & \Circle \\[0.2cm]

\cite{villain2022_sysj} & Spectrogram & PCA & IEEE 802.11 & OFDM & AWGN & \Circle & \CIRCLE & \Circle  & \LEFTcircle & \Circle \\[0.2cm]
 
\cite{alhazbi2023_ccnc}    & I/Q & CNN & - & BPSK & BPSK & \CIRCLE & \CIRCLE & \Circle & \LEFTcircle & \Circle\\
\cite{sciancalepore2024_iotj}   & I/Q & AE & - & BPSK & \makecell{AWGN, \\ BPSK} & \CIRCLE & \CIRCLE & \CIRCLE & \LEFTcircle & \Circle\\ [0.2cm]
\hline
\makecell{This \\ paper}   & I/Q & CNN, AE & IEEE 802.11 & \makecell{OFDM w/ \\ B/QPSK, \\ 16/64-QAM} & \makecell{AWGN, \\ BPSK} & \CIRCLE & \CIRCLE & \CIRCLE & \CIRCLE & \CIRCLE\\

 \hline
\end{tabular}
\label{tab:related}
\end{table*}
Jamming detection is usually performed by analyzing one or more communication link metrics, e.g., the \ac{PDR}, the \ac{PLR} and \ac{PRR}.
Some approaches detect jamming through the analysis of \ac{PHY}-layer metrics, such as the \ac{RSS}~\cite{saxena2022_pmc}, while others consider the analysis of raw signals, represented in different ways. For example, the authors in~\cite{toma2020_tccn} pre-process IQ samples with a \ac{ST}, while the authors in~\cite{li2022_access} extract the spectrogram from the raw signal, and the authors in~\cite{gecgel2019_wiseml} use the \ac{STFT} and the \ac{CGDWT} to extract more information from the raw signal. Others use the \ac{CER}, which is derived from the chip-to-symbol conversion process \cite{spuhler2014_twc}, the Residual Energy Decay Rate \cite{furqan2020_jwcn} or the Logarithmic Received Energy \cite{zhang2023_tccn}. In contrast to using only one feature, multiple heuristic signal processing features can also be used, as discussed in~\cite{kihei2021_wfiot}. 

The recent increased popularity of \ac{AI} has contributed to an increase in the usage of \ac{ML}- and \ac{DL}-based approaches for the analysis of the chosen link metric(s).
We can notice a great variety in the usage discriminative \ac{AI}-based tools, including Random Forests \cite{punal2014_wowmom, li2022_access, arjoune2020_icoin}, \ac{K-NN} \cite{kihei2021_wfiot}, \ac{CNN} \cite{li2022_access, alhazbi2023_ccnc}, Deep \ac{CNN}, Deep Recurrent Neural Networks \cite{gecgel2019_wiseml} and Sparse \acp{AE} \cite{sciancalepore2024_iotj}. A generative \ac{ML} model, \ac{C-GAN} is used in \cite{toma2020_tccn}. There are also attempts to use non-\ac{ML} based approaches, such as the \ac{HHT} \cite{spuhler2014_twc} and \acp{HMM} \cite{zhang2023_tccn}. Some recent papers, such as~\cite{savolainen2024_icl} and~\cite{chitauro2024_iswcs}, considered the problem of jamming classification at the \ac{PHY} layer, especially in the context of satellite signals. However, such papers aim to detect and classify jamming, not to identify it when its power is low compared to the communication signal. Similarly, the authors in~\cite{smailes2024_arxiv} investigated the resilience of satellite fingerprinting schemes to jamming while not investigating jamming detection.

In this paper, we focus on the detection of \emph{weak} jamming, i.e., detecting jamming in a situation where the level of the received jammer power at a target is not enough to shut down the communication link entirely but only to affect it slightly. In our earlier contributions, we denoted such a scenario as a \emph{Low-BER Regime} since it translates into a low (non-zero) \ac{BER} at the target. We notice that weak-jamming detection is relevant for static and mobile networks. Mobile devices, such as with \acp{UAV} and autonomous vehicles, could detect jamming while moving toward the jammer's location. In these contexts, weak-jamming detection enables the pilot and the \ac{GCS} to keep control of the vehicle rather than losing control and relying on a pre-programmed action to navigate safely. \textcolor{black}{Although some works in the literature consider tests for jamming under mobility, e.g., \cite{punal2014_wowmom} and \cite{spuhler2014_twc}, to the best of our knowledge, only a few papers delved into the detection of weak and low-power jamming and proposed solutions to this problem. The pioneering contribution by Garnaev et al. in~\cite{garnaev2016_tsp} only provided numerical analysis, while the contributions by Boganoski et al.~\cite{bogdanoski2014_conf} and Venkata et al.~\cite{venkata2021_wlett} used simulations to prove the feasibility of detecting low-power jamming. The authors in~\cite{venkata2021_wlett} used SVM and Random Forest classifiers to detect jamming, identifying early the potential of \ac{ML} to address this problem. However, they considered the scenario of vehicular networks and their proposed methods took significant time (in the order of seconds), not being suitable to highly dynamic conditions. More recently, Villain et al.~\cite{villain2022_sysj} were the first to use real-world data collected in an office environment to show that, using Principal Component Analysis (PCA), it is possible to detect low-power jamming. Such works mostly focus on spectral differences, while providing limited quantitative results across various jamming detection parameters. Indeed, they do not consider the impact of the distance between transmitter, receiver and jammer, and not even mobile deployments. They also consider an external device specifically meant for jamming detection. }
A step further in this direction is offered by our recent contributions in \cite{alhazbi2023_ccnc} and \cite{sciancalepore2024_iotj}, which cover many different properties of low-power-jamming detection, including the impact of the distance between the jammer and the communication link. They are also the only contributions releasing the data and the code used for their experiments, encouraging reproducibility. Although using real-world data, those contributions consider simple communication links, not aligning with real-world operational environments, and test their solutions indoors. In this paper, we advance further the state of the art on weak-jamming detection by (i) extending the methodology defined in~\cite{alhazbi2023_ccnc} and \cite{sciancalepore2024_iotj} for more complex modulation schemes; (ii) using an actual communication technology used by many wireless deployments, including drones and connected vehicles, i.e., WiFi, (iii) evaluating weak-jamming detection with various modulation schemes used in the real-world WiFi, i.e., \ac{OFDM} with \ac{BPSK}, \ac{QPSK}, 16-\ac{QAM} and 64-\ac{QAM}, and (iv) performing extensive tests outdoors, considering various signal propagation conditions (indoors, outdoor garden and outdoor field). \textcolor{black}{Moreover, compared to early works on low-power jamming detection, this paper makes significant advancements: (i) extensive real-world data collection, (ii) extensive experiments and results considering various jamming detection parameters, (iii) use of state-of-the-art \ac{DL}-based solutions, and (iv) very limited collection and inference time, applicable to real-time scenarios.}
\textcolor{black}{Finally, we highlight that the image generation technique used in our paper shares similarities with the one recently used for Radio Frequency Fingerprinting (RFF), e.g., in~\cite{oligeri2024_sac},~\cite{alhazbi2023_acsac}, and~\cite{papangelo2023commag}. The proposal in~\cite{oligeri2024_sac} is meant to detect spoofing attacks to satellite constellation through the analysis of the physical-layer information available at the receiver. To this aim, the authors generated images from \ac{IQ} samples and applied classification to detect spoofing attacks. The authors in~\cite{alhazbi2023_acsac} showed that the power-cycles of the radios affect the performance of RFF solutions, and that techniques based on images can improve performance compared to analying raw physical-layer data. The authors in~\cite{papangelo2023commag} show that Adversarial Machine Learning techniques can be used to improve the robustness of image-based RFF techniques, so avoiding synthetic attacks via imitation of the features of the physical-layer signals at a receiver. 
However, using such image generation technique for the detection of weak jamming entails different choices and analysis methods, not affecting the novelty and aim of our contribution.}

\section{Scenario and Adversary Model}
\label{sec:scenario_adv_model}
\begin{figure}
\centering
    \subfloat[Mobility]{\includegraphics[width=0.494\columnwidth]{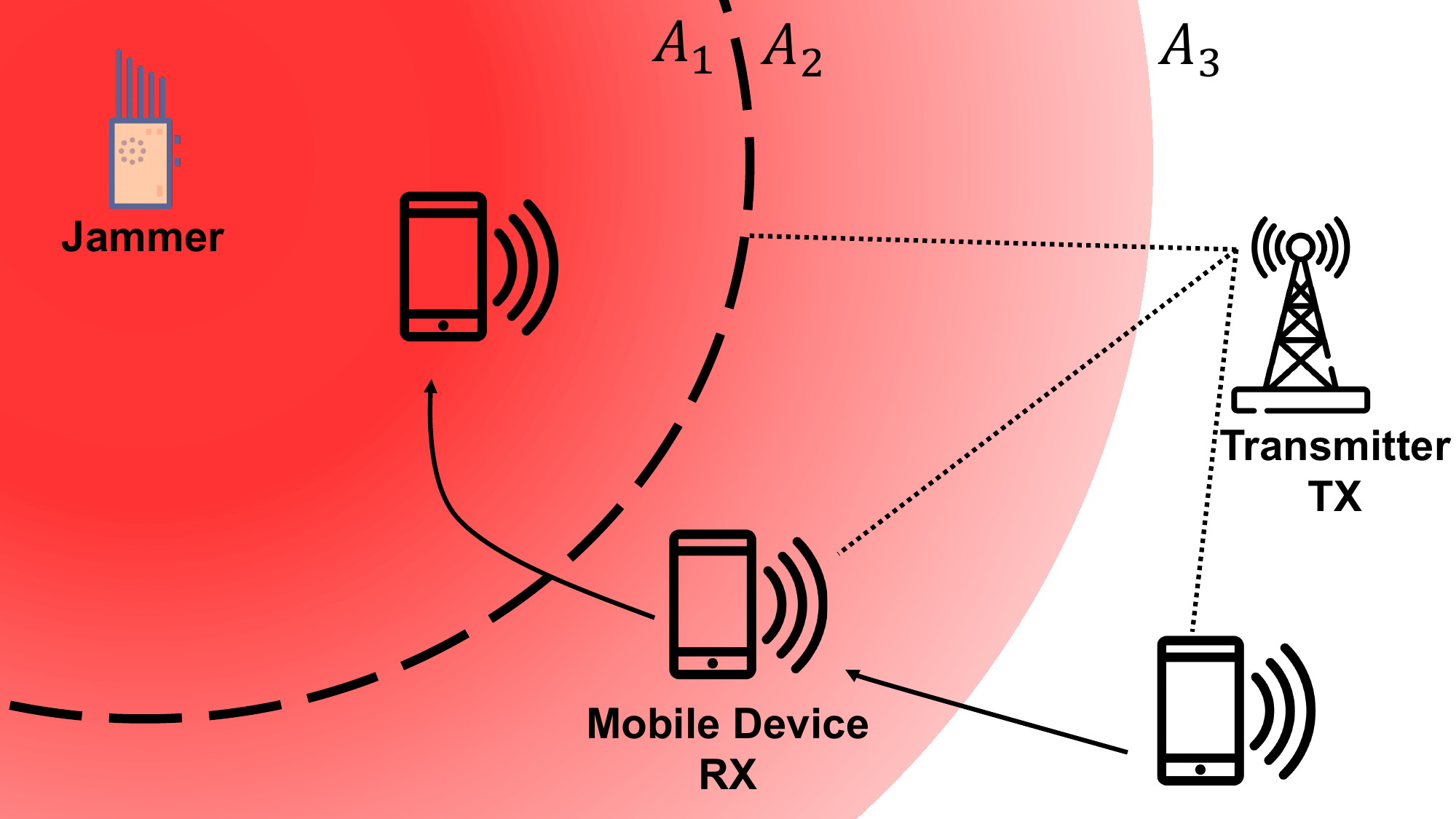}}
    \hfill
    \subfloat[IoT]{\includegraphics[width=0.493\columnwidth]{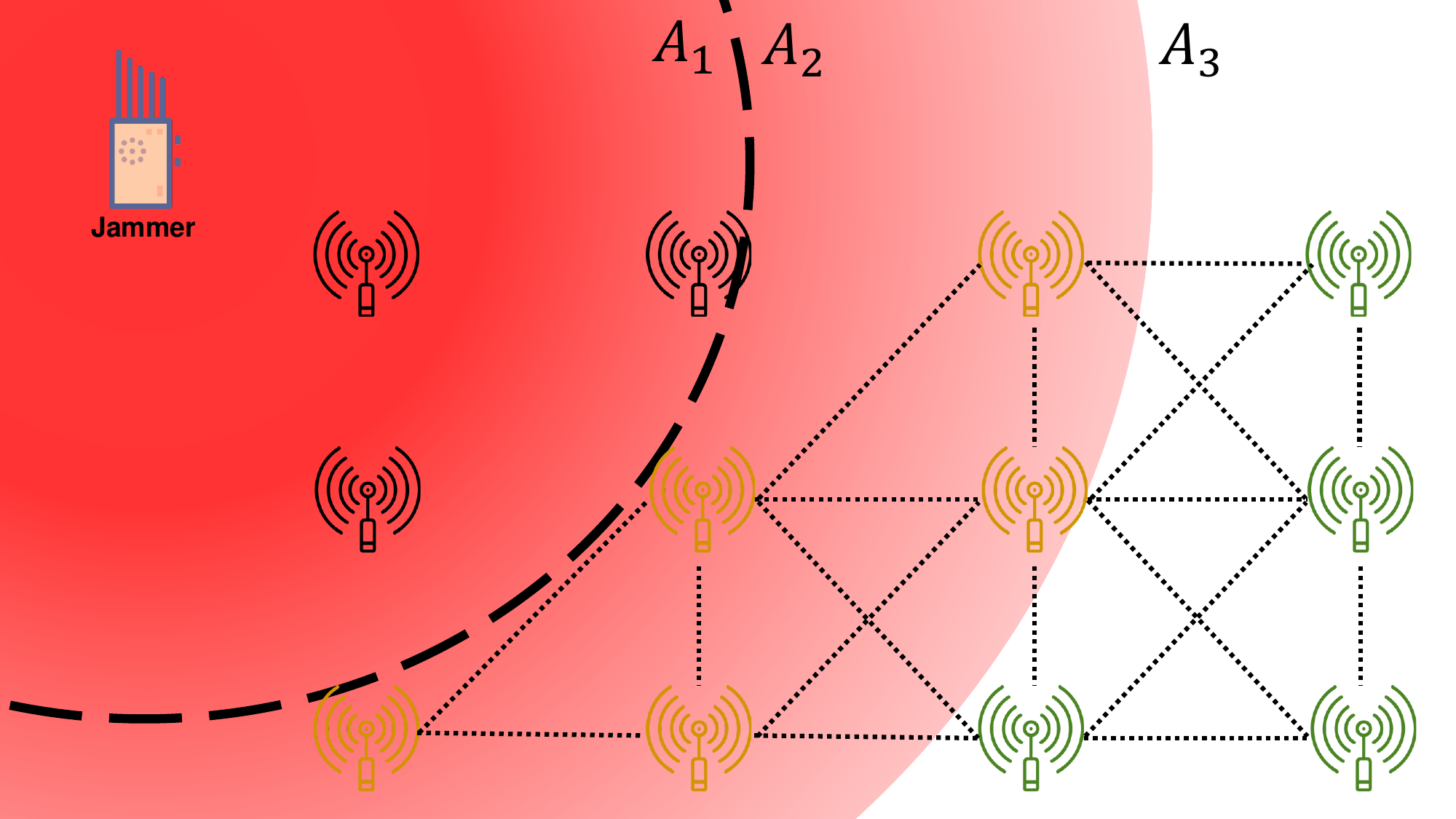}}
    \caption{System and Adversary Model. Our solution applies to two reference scenarios. (a) A receiver (RX) moves from area $A_3$ to area $A_2$ and eventually to $A_1$. RX can receive messages from TX when in $A_2$ and $A_3$, while reception is not possible in $A_1$ due to jamming. (b) A dense network is partially jammed (black devices in $A1$) while (yellow and green) devices in $A_2$ and $A_3$ can communicate.}
    \label{fig:scenario}
\end{figure}

Figure~\ref{fig:scenario} depicts the reference scenarios and the adversary model considered in this work. 

We consider two reference scenarios, i.e., Fig.~\ref{fig:scenario}(a), representing a mobile receiver (RX) moving towards a jammed area while communicating with a remote transmitter (TX), and Fig.~\ref{fig:scenario}(b), showing a partially jammed dense network of \ac{IoT} devices, simultaneously acting as TX and RX. For both scenarios, the presence of the jammer splits the playground into three areas: $A_1$, $A_2$, and $A_3$, respectively. No message reception is possible in the area $A_1$ as the jamming signal overcomes the power of the legitimate signal of the transmitter. In contrast, communication is possible in areas $A_3$ and $A_2$, but in $A_2$, the devices experience both the legitimate signal and the jamming. In particular, devices deployed in $A_2$ can infer the presence of a jammer through the analysis of the received signal at the \ac{PHY} layer.

Without loss of generality, we consider that the RX (\ac{IoT}) device features an omnidirectional transceiver antenna, in line with the equipment onboard regular wireless devices. We consider the IEEE 802.11g communication technology based on its common usage in several \ac{COTS} \ac{IoT} devices. We recall that IEEE 802.11g works in the 2.4GHz ISM band and supports the following modulation schemes: \ac{OFDM} with \ac{BPSK}, \ac{QPSK}, 16-\ac{QAM} and 64-\ac{QAM}.

{\bf Adversary Model.} We consider an adversary $\mathcal{A}$ equipped with an omnidirectional antenna that emits a jamming signal to disrupt the radio communication nearby (area $A_1$). The jamming signal can be noise, i.e., \ac{AWGN}, or deceptive, i.e., adopting the same modulation scheme used by the legitimate communication link. We notice that deceptive jamming assumes some level of prior knowledge about the communication link. Furthermore, we consider an adversary that constantly jams at the highest possible transmission power. The usage of the highest transmission power is consistent with the general consideration that the jammer wants to prevent any communication in the nearby area as much as possible while not knowing the specific location of the RX and the TX. We also consider a jammer unaware of the sampling rate used by the legitimate communication link. Thus, it transmits the jamming signal with the highest possible sampling rate as a function of the available hardware. Note that weak-jamming occurs even when the jammer emits jamming with high power: indeed, if the receiver is sufficiently farther away, the level of the injected noise is not enough to jam the channel completely, creating a weak-jamming scenario.

Finally, we highlight that area $A2$ is characterized by a bit error rate below $1\%$. This is the area that we will focus on in this work. In fact, in this region, the RX (IoT) wireless device can (potentially) detect the presence of the jammer and has a reliable communication link with the TX, thus allowing awareness and informed decisions on the following actions. Detecting weak jamming signals increases situational awareness and allows the TX to take action accordingly without having to rely on pre-defined backup steps pre-programmed on the RX device.

\section{Detection of Weak Jamming Signals}
\label{sec:methodology}
In this section, we introduce our enhanced methodology for weak-jamming detection. Overall, we can identify two main phases, i.e., \emph{Image Generation} (Sec.~\ref{sec:image_generation}) and \emph{Jamming Detection} (Sec.~\ref{sec:jam_detect}). Tab. \ref{tab:notation} summarizes the main notation used throughout the section.

\begin{table}
\caption{Notation and brief description.}
    \label{tab:notation}
    \centering
    \scriptsize
    \begin{tabular}{c|c}
       \textbf{Notation} & \textbf{Description} \\
       \hline
        $P, Q$ & Image dimensions (Width, Height). \\
        $n$ & Number of \ac{IQ} samples per image. \\
        $\tau$ & Threshold value for the \ac{AE}. \\
        $MSE_{train}$ & \makecell{MSE value from reconstructing \\the trainset.} \\
    \end{tabular}
\end{table}

\subsection{Image Generation}
\label{sec:image_generation}

\begin{figure}
    \centering
    \includegraphics[width=\columnwidth]{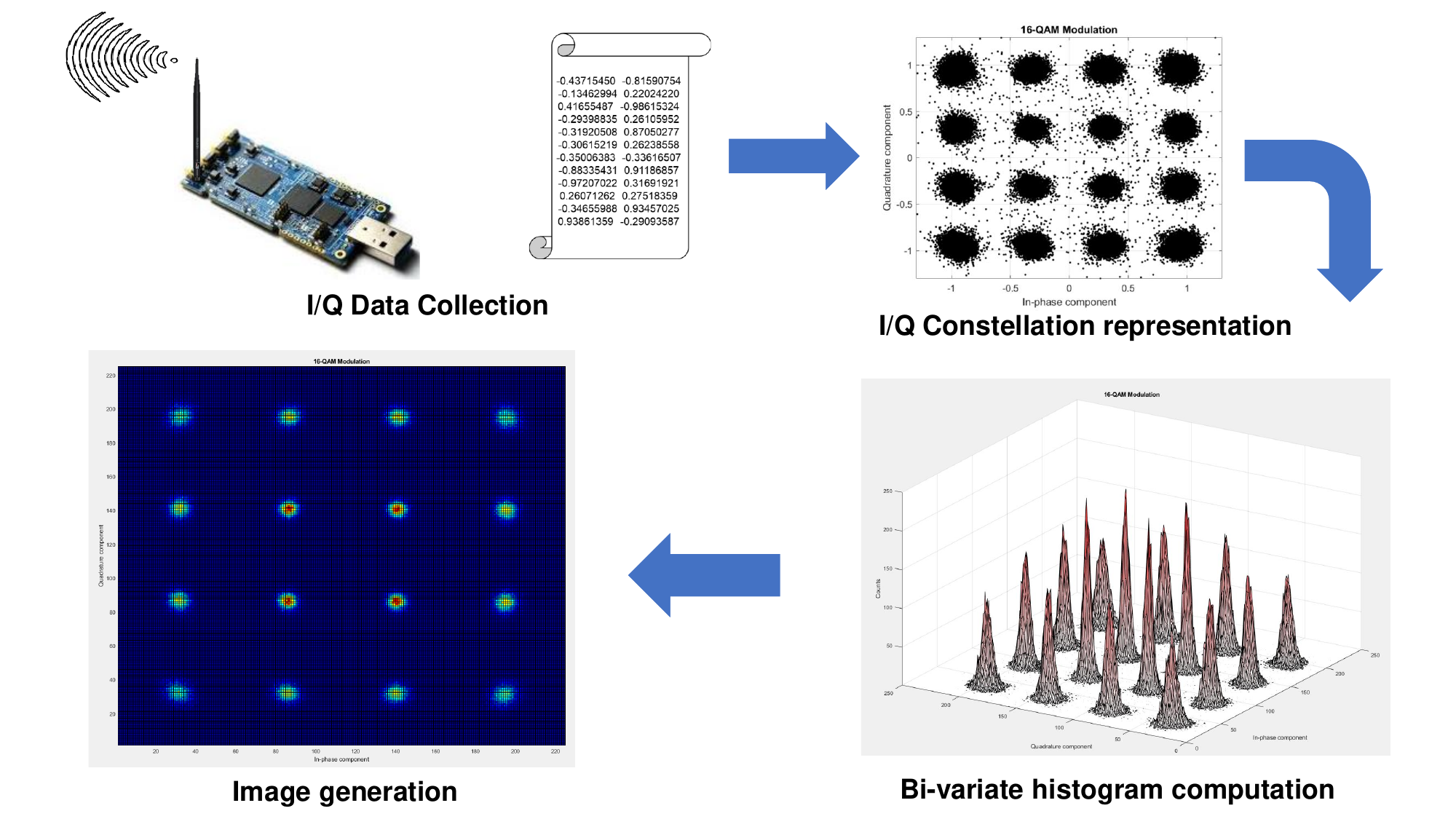}
    \caption{\textcolor{black}{We collect the IQ data through a \ac{SDR}, then divide them into chunks of $n$ samples per image, represent them in a constellation diagram, compute a bi-variate histogram, and store it as an image.}
    }
    \label{fig:Image_Gen_Process}
\end{figure}

Figure \ref{fig:Image_Gen_Process} presents an overview of the adopted image generation pipeline.
\textcolor{black}{We collect \ac{IQ} samples using a device capable of acquiring \ac{PHY}-layer wireless channel information, e.g., an \ac{SDR}. Next, we divide these samples into subsets of $n$ samples, denoted as the \emph{number of samples per image} (see Sec.~\ref{sec:results} for details on the impact of this parameter on jamming detection).} Then, we represent each subset of $n$ \ac{IQ} samples through the \ac{IQ} constellation diagram, with the \emph{In-phase} component on the x-axis and the \emph{Quadrature} component on the y-axis. We highlight that our image processing technique differs from the previously proposed approach, being more suitable for higher-order modulation schemes. Previous solutions~\cite{alhazbi2023_ccnc} and~\cite{sciancalepore2024_iotj} considered only one of the two clouds in the image. In contrast, we consider the \ac{IQ} constellation with x-axis and y-axis limits of $[-2,2]$ and divide it into a $P \times Q$ grid. \textcolor{black}{Note that, at the receiver, we normalize the value of the IQ samples, thus ensuring that their values do not fall outside of the considered range.
We apply dataset-level normalization, i.e., we compute the maximum values of raw I and Q ($I_{MAX}$, $Q_{MAX}$), and we normalize the samples as $I = \frac{I}{I_{MAX}}$ and  $Q = \frac{Q}{Q_{MAX}}$. }
We use the mentioned grid to generate a bi-variate histogram by counting how many \ac{IQ} samples fall into a specific tile of the grid. The bins in the histogram are considered to be the value of the image pixels. Lastly, the output of the histogram is saved as a $P \times Q$ grayscale image, where $P$ and $Q$ are selected based on the input requirements of the classifier used later in the jamming detection phase. In Sec.~\ref{sec:results}, we experimentally compare the proposed image generation method with the technique used in~\cite{alhazbi2023_ccnc} and~\cite{sciancalepore2024_iotj} and evaluate the pros and cons.

\subsection{Jamming Detection}
\label{sec:jam_detect}

We consider two different approaches, i.e., binary classification using \acp{CNN}, as in~\cite{alhazbi2023_ccnc}, and anomaly detection using \emph{sparse \acp{AE}}, as in~\cite{sciancalepore2024_iotj}. While the former requires the knowledge of samples affected by jamming, the latter can be deployed in scenarios where jamming has never been observed during the training of the model, reducing the training overhead.

\begin{figure}[h]
    \centering
    \includegraphics[width=\columnwidth]{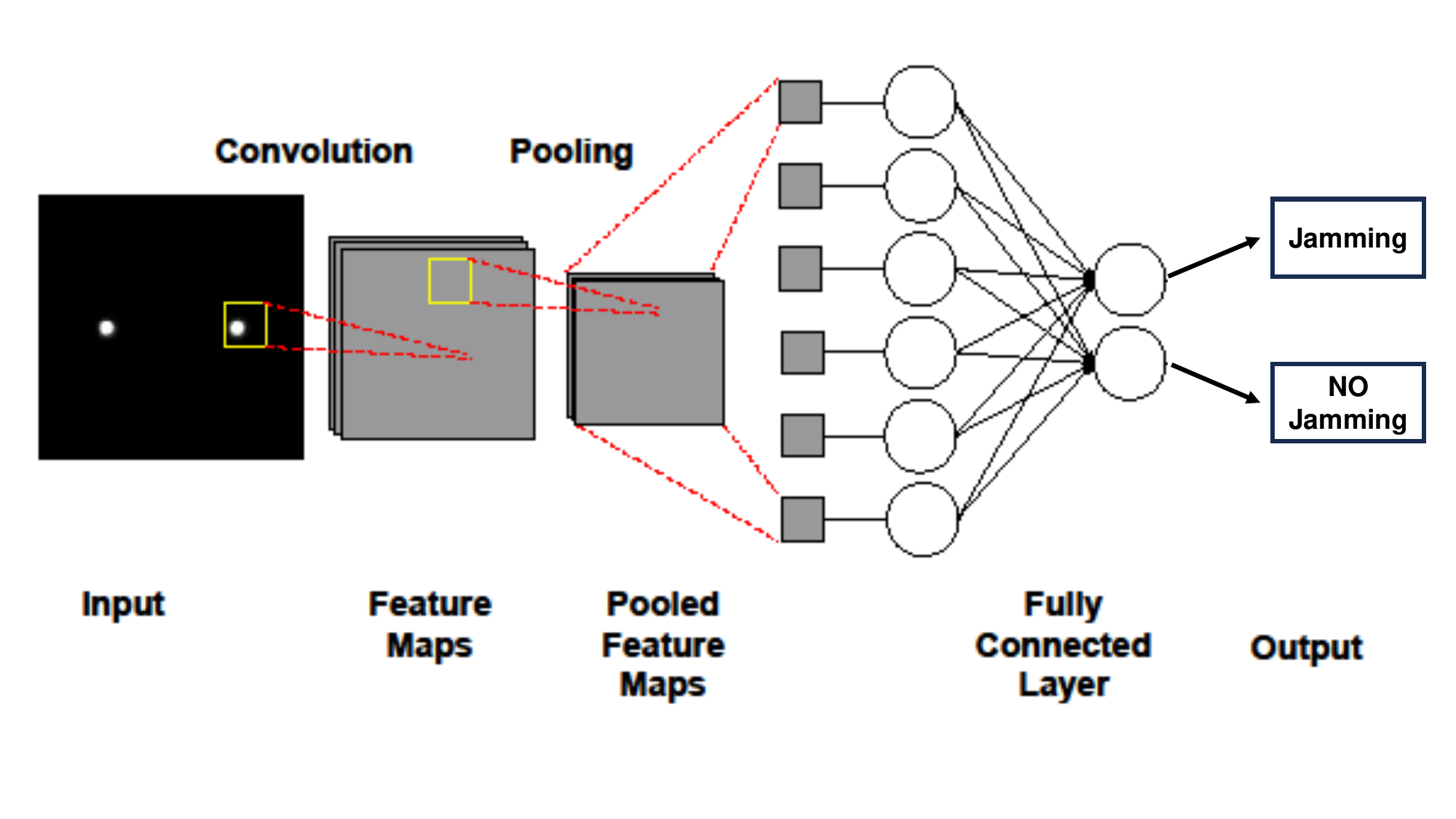}
    \caption{Visual representation of a the \ac{CNN} used for binary jamming detection.}
    \label{fig:CNN}
    \end{figure}

\textbf{Binary Classification via \acp{CNN}.} This approach detects jamming based on a binary classifier. More in detail, we use a \ac{CNN} with 18 residual layers, i.e., ResNet-18~\cite{ResNet-18} which is pre-trained on the ImageNet dataset~\cite{russakovsky2015}. We adapt the \ac{CNN} to have two output classes to distinguish between jamming and no jamming. We present a simplified visual overview of the architecture in Fig.~\ref{fig:CNN}. 
We consider a training process characterized by 30 epochs and a mini-batch size of 35 elements. We use images generated from \ac{IQ} samples captured with no jamming affecting the communication link (\emph{NO JAM} images) and an equal number of images generated from \ac{IQ} samples captured with jamming (referred to as \emph{JAM} images) for the training process. The final output of the algorithm is the decision between jamming or no jamming. \textcolor{black}{For hyper-parameters tuning, we referred to the same configuration used in~\cite{sciancalepore2024_iotj}.}

\begin{figure}[h]
    \centering
    \includegraphics[width=\columnwidth]{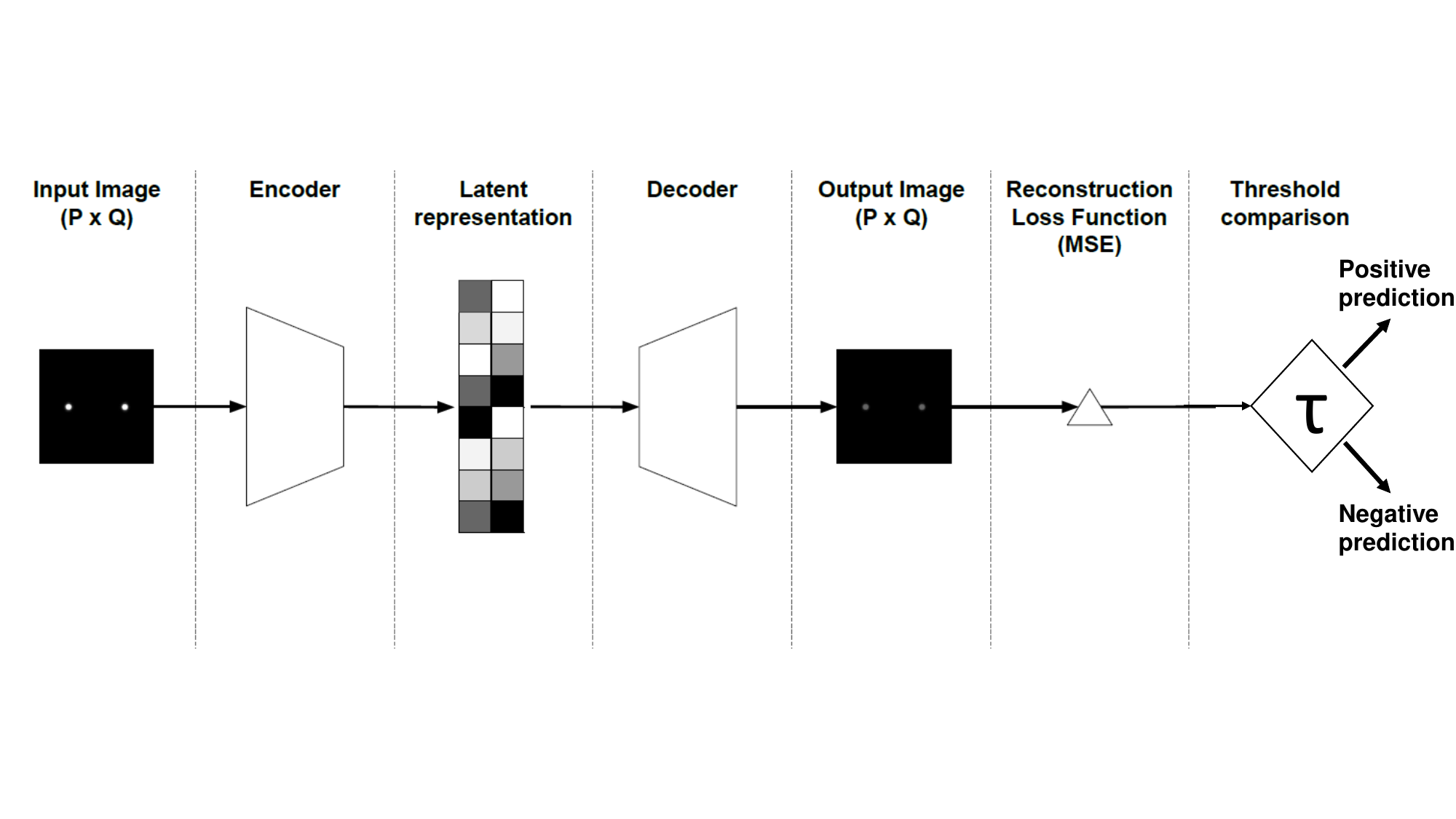}
    \caption{Visual representation of the \ac{AE}-based jamming detection architecture.}
    \label{fig:Autoencoder_Arch}
\end{figure}

\textbf{Anomaly Detection via \acp{AE}.} This approach detects jamming by identifying anomalies in the received signal, using sparse \acp{AE}. Figure~\ref{fig:Autoencoder_Arch} shows the adopted \ac{AE} architecture.
We consider the generated images as described in Sec.~\ref{sec:image_generation} as input of an encoder using a logarithmic sigmoid function with $K=16$ neurons, obtaining a compressed latent representation of the input image consisting of $K=16$ dimensions. Then, we feed the latent representation vectors to a decoder using a linear decoder transfer function with a total number of $J = 50,176$ neurons, as described in~\cite{sciancalepore2024_iotj}. We use two hidden layers and the sparsity regularization technique in line with the methodology provided in~\cite{sciancalepore2024_iotj}. The output is an image of the same size as the input image ($P\times Q$). We convert this image into a matrix and compare it with the input through the \ac{MSE} loss function. 
This approach entails \emph{training} and \emph{testing}. For the training process, we use images generated from \ac{IQ} samples captured with no jamming affecting the communication link, generating a set of \ac{MSE} values representing the regular behavior of the link. We use these values to calculate the threshold $\tau$ to differentiate images generated from non-jamming or jamming samples. An image with $MSE < \tau$ is predicted to be generated from no jamming samples, while an image with $MSE > \tau$ is predicted to be generated from jammed samples. For computing $\tau$, we adopt Eq.~\ref{eq:Threshold_Formula}, as proposed by the authors in \cite{erfani2016_prec}.
\begin{equation}
    \label{eq:Threshold_Formula}
    \tau = E(MSE_{train}) + 3.5 \cdot \sigma(MSE_{train}),
\end{equation} 
where $E(\circ)$ and $\sigma(\circ)$ represent the mean and the standard deviation, respectively.

\begin{figure}
    \centering
    \includegraphics[width=\columnwidth,angle = 0,trim = 30mm 80mm 30mm 90mm]{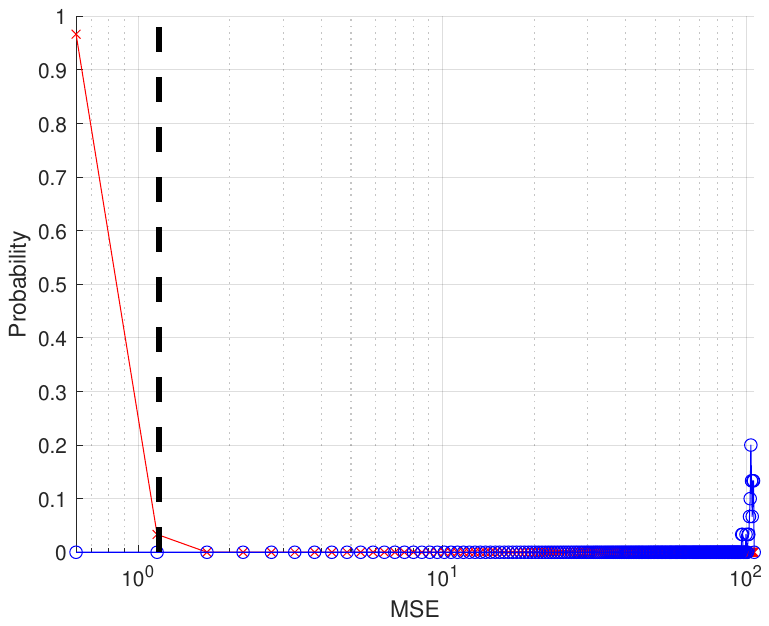}
    \caption{Sample distribution of the \ac{MSE} values from our dataset, with \emph{NO JAM} images (red), \emph{JAM} (blue) images, and threshold value (dashed line).}
    \label{fig:MSE_Distr}
\end{figure}

An image generated from unjammed samples is characterized by lower \ac{MSE} values. In contrast, an image generated from jammed samples results in a higher \ac{MSE} value. We show a sample distribution of the \acp{MSE} in Fig.~\ref{fig:MSE_Distr}.
\textcolor{black}{Finally, note that we select the same values for the hyperparameters of the \ac{AE} and \acp{CNN} as the approach presented in~\cite{sciancalepore2024_iotj}, i.e., a hidden size value of $16$, a sparsity regularization term of $0.5$, an L2-regularization term of $0.01$, the Logistic Sigmoid (logsig) function as the encoder transfer function and $250$ epochs for the training process.}

\section{Data Collection and Experiment Settings}
\label{sec:data_experiments}
In this section, we describe the methodology used for data collection, including the hardware and software details of our measurement setup (Sec.~\ref{sec:hw_sw}) and the settings used for the experiments (Sec.~\ref{sec:settings}).

\subsection{Hardware and Software}
\label{sec:hw_sw}
{\bf Hardware.} We consider the \acp{SDR} provided by Lime Microsystems, i.e., three LimeSDR USB \cite{limesdr_info} devices and one LimeSDR Mini v2.0 \cite{limesdr_mini_2.0}, equipped with a WiFi-capable antenna and an up-to-date firmware. The LimeSDR USB has a frequency range of $100$~kHz up to $3.8$~GHz, a maximum bandwidth of $61.44$~MHz, and a transmit power of up to $10$~dBm. The LimeSDR Mini V2.0 has a smaller frequency range of $10$~MHz up to $3.5$~GHz, $40$~MHz of maximum bandwidth, and a maximum transmitting power of $10$~dBm.
We use two LimeSDR USB devices to establish an \ac{RF} communication link, thus representing the \ac{GCS} and \ac{UAV}. Another LimeSDR USB and the LimeSDR Mini v2.0 are used as jammers to disrupt the communication link. We use two general-purpose laptops to control the \acp{SDR}. For data analysis, we upload the captured \ac{IQ} samples to a centralized computing facility, i.e., the \ac{HPC} cluster hosted at TU/e, Eindhoven, Netherlands. It provides an AMD EPYC 7313 CPU running at 3.00 GHz, 16 GB of RAM, and an NVIDIA A30 GPU with 24GB of RAM.

{\bf Software.} We use GNURadio version 3.10~\cite{gnuradio-3.10} and the block \emph{gr-limesdr}~\cite{gr-limesdr} to interact with the \acp{SDR}. The block diagram used for this work builds on top of the publicly available out-of-tree module \emph{gr-ieee-802-11}, available at~\cite{gr-ieee802-11}, allowing the transmission and reception of IEEE 802.11g packets through \acp{SDR}, with a few relevant modifications. Specifically, to make the block diagram work correctly on our hardware, we added two blocks to the receiver chain, i.e., an \emph{FFT Filter} and \emph{DC Spike Remover}. Such modifications filter out the DC offset affecting the used hardware and make it possible to correctly receive \ac{IQ} samples. We also save the \ac{IQ} data corresponding to correctly received WiFi packets to a file, in the \emph{fc16} format, useful for post-processing according to the techniques described in Sec.~\ref{sec:methodology}. We also modify the block diagram of the WiFi transmitter by adding a block \emph{Multiply Const} just before the block \emph{LimeSDR Sink}, with an input value of $0.5$. This modification is meant to reduce the amplitude of the resulting signal and make it transmittable by the LimeSDR without distortions. \textcolor{black}{On the receiving LimeSDR, we collect the \ac{IQ} samples corresponding to the whole IEEE 802.11g packet, including both the header and payload. We collect IQ samples from the wireless channel using a limited sample rate of $5 \cdot 10^6$~\ac{sps}, as the link becomes unreliable at higher sample rates (see Sec.~\ref{sec:discussion} for a discussion of the impact of this choice on our results). We consider a transmission rate of one IEEE 802.11g frame every $115$~ms, bounded by the capabilities of our hardware. The payload consists of two \textit{x} characters. For the jammer, we use the blocks \emph{Gaussian Noise Source} to generate samples, \emph{Multiply Const} with input value $0.5$ to avoid RF signal distortion, and \emph{LimeSDR Sink} to transmit the samples to the hardware. As for the implementation of the deceptive jamming, we consider the same flow diagram used for the transmitter. Finally, note that we set the carrier frequency for all \acp{SDR} to $f_c = 2,484$ MHz, equal to channel 14 as described by IEEE 802.11g~\cite{IEEE802.11g}.} 

\subsection{Measurement Settings}
\label{sec:settings}

\textcolor{black}{Table~\ref{tab:dataset} provides the details of the dataset collected for this work. It describes, for each environment, the tested control variable, experiment-specific settings, tested jamming conditions, collected IQ samples, dataset size, and Symbol Error Rate (SER). Note that we collected over 6.6 GB of data and 843M+ IQ data. To the best of our knowledge, this is the largest dataset for weak jamming scenarios including both indoor and outdoor measurements.}

\begin{table*}[!t]
\caption{\textcolor{black}{Details of the dataset collected as part of our experiments.}
}
\label{tab:dataset}
\centering
\color{black}
\scriptsize
\begin{tabular}{c|l|l|l|l|l|l}
\textbf{Env.}               & \textbf{Control Variable}             & \textbf{\begin{tabular}[c]{@{}l@{}}Jamming \\ Conditions\end{tabular}} & \textbf{Experiment-specific Settings} & \multicolumn{1}{c|}{\textbf{\begin{tabular}[c]{@{}c@{}}Collected \\ IQ Samples\end{tabular}}} & \textbf{\begin{tabular}[c]{@{}l@{}}Dataset \\ Size {[}MB{]}\end{tabular}} & \textbf{SER} \\ \hline
\multirow{11}{*}{E1}        & \multirow{2}{*}{Jammer Distance}      & No Jam                      &                                       & 80740310                      & 61,6                           & 0,0583       \\ \cline{3-7} 
                            &                                       & Jam                         & 3, 5, 7,10,13,16,19,21,23,25 m        & 43729434                      & 309,35                         & 0,0663       \\ \cline{2-7} 
                            & \multirow{2}{*}{Receiver Distance}    & No Jam                      & 5,7,10,13,16,19,21,23,25,27 m         & 7986299                       & 633,52                         & 0,0008       \\ \cline{3-7} 
                            &                                       & Jam                         & 5,7,10,13,16,19,21,23,25,27 m         & 43950176                      & 343,37                         & 0,0107       \\ \cline{2-7} 
                            & \multirow{2}{*}{Transmitter Distance} & No Jam                      & 5,7,10,13,16,19,21,23,25,27 m         & 79805862                      & 623,49                         & 0,0127       \\ \cline{3-7} 
                            &                                       & Jam                         & 5,7,10,13,16,19,21,23,25,27 m         & 47540014                      & 371,43                         & 0,0330       \\ \cline{2-7} 
                            & Reference Setup                       & No Jam                      &  Location 0 (Ref. Setup)                                 & 8074310                       & 63,08                          & 0,0583       \\ \cline{2-7} 
                            & Jammer   Gain                         & Jam                         & 6,12,14,16,18,20                      & 24160008                      & 188,75                         & 0,0821       \\ \cline{2-7} 
                            & Jammer   Location                     & Jam                         & 1, 2, 3                               & 12112545                      & 94,63                          & 0,0763       \\ \cline{2-7} 
                            & Jamming   Oversampling Ratio          & Jam                         & 1, 2, 3, 4                            & 16150765                      & 126,18                         & 0,0788       \\ \cline{2-7} 
                            & Jammer   Type                         & Jam                         & AWGN, Deceptive                       & 12185053                      & 94,5                           & 0,4287       \\ \hline
\multirow{15}{*}{E2}        & Reference Setup                       & No Jam                      & 0, 3 m                                & 16769451                      & 131,01                         & 0,0542       \\ \cline{2-7} 
                            & Jammer   Distance                     & Jam                         & 3, 5, 7,10,13,16,19,21,23,25, 27   m  & 47542742                      & 371,42                         & 0,0823       \\ \cline{2-7} 
                            & \multirow{2}{*}{Receiver Distance}    & No Jam                      & 5, 7, 10, 13, 16, 19, 21 m            & 59417153                      & 464,17                         & 0,0526       \\ \cline{3-7} 
                            &                                       & Jam                         & 5, 7, 10, 13, 16, 19, 21 m            & 30666679                      & 239,58                         & 0,1899       \\ \cline{2-7} 
                            & \multirow{2}{*}{Transmitter Distance} & No Jam                      & 5, 7, 10, 13, 16, 19, 21 m            & 59417153                      & 464,17                         & 0,1256       \\ \cline{3-7} 
                            &                                       & Jam                         & 5, 7, 10, 13, 16, 19, 21 m            & 30389065                      & 237,4                          & 0,0917       \\ \cline{2-7} 
                            & Reference   Setup                     & No Jam                      &                                       & 7951440                       & 62,12                          & 0,0302       \\ \cline{2-7} 
                            & Jammer   Gain                         & Jam                         & 24,26,28,30,32                        & 16312240                      & 155,41                         & 0,0244       \\ \cline{2-7} 
                            & Jammer   Location                     & Jam                         & 1, 2, 3                               & 12119906                      & 94,68                          & 0,0838       \\ \cline{2-7} 
                            & Jamming   Oversampling Ratio          & Jam                         & 1, 2, 3, 4                            & 16160104                      & 126,25                         & 0,0719       \\ \cline{2-7} 
                            & Jammer   Type                         & Jam                         & AWGN, Deceptive                       & 11887393                      & 92,17                          & 0,4408       \\ \cline{2-7} 
                            & Jammer   Hardware                     & Jam                         & LimeSDR Mini, LimeSDR USB             & 8080602                       & 63,14                          & 0,0652       \\ \cline{2-7} 
                            & Modulation   Type                     & Jam                         & BPSK, QPSK, 16-QAM, 64-QAM            & 0                             & 79,79                          & 0,0000       \\ \cline{2-7} 
                            & \multirow{2}{*}{RX Mobility}          & No Jam                      & 0, 1, 2                               & 27696974                      & 216,38                         & 0,0026       \\ \cline{3-7} 
                            &                                       & Jam                         & 0, 1, 2                               & 14032334                      & 109,64                         & 0,0158       \\ \hline
\multirow{7}{*}{E3}         & Reference Setup                       & No Jam                      &                                       & 7957589                       & 60,7                           & 0,0392       \\ \cline{2-7} 
                            & Jammer   Gain                         & Jam                         & 6,12,14,16,18,20                      & 19897284                      & 152                            & 0,0391       \\ \cline{2-7} 
                            & Jammer Location                       & Jam                         & 1, 2, 3                               & 12116831                      & 92,4                           & 0,0744       \\ \cline{2-7} 
                            & Jamming Oversampling Ratio            & Jam                         & 1, 2, 3, 4                            & 16163242                      & 123,3                          & 0,0668       \\ \cline{2-7} 
                            & Jammer Type                           & Jam                         & AWGN, Deceptive                       & 12200389                      & 94,61                          & 0,0900       \\ \cline{2-7} 
                            & Jammer Hardware                       & Jam                         & LimeSDR Mini, LimeSDR USB             & 8077193                       & 63,1                           & 0,1024       \\ \cline{2-7} 
                            & Transmitter Gain                      & Jam                         & 42, 47, 52, 57                        & 31840829                      & 248,75                         & 0,0697       \\ \hline
\multicolumn{1}{l|}{Total} &                                       &                             &                                       & 843131369                     & 6652,09                        &              \\ 
\end{tabular}
\end{table*}

We consider three different measurement environments characterized by an increased amount of background noise, namely, Outdoor Low-Multipath (E1), Outdoor Multipath (E2), and Indoor (E3). 

\begin{figure}[h]
    \centering 
    \includegraphics[width=.8\columnwidth]{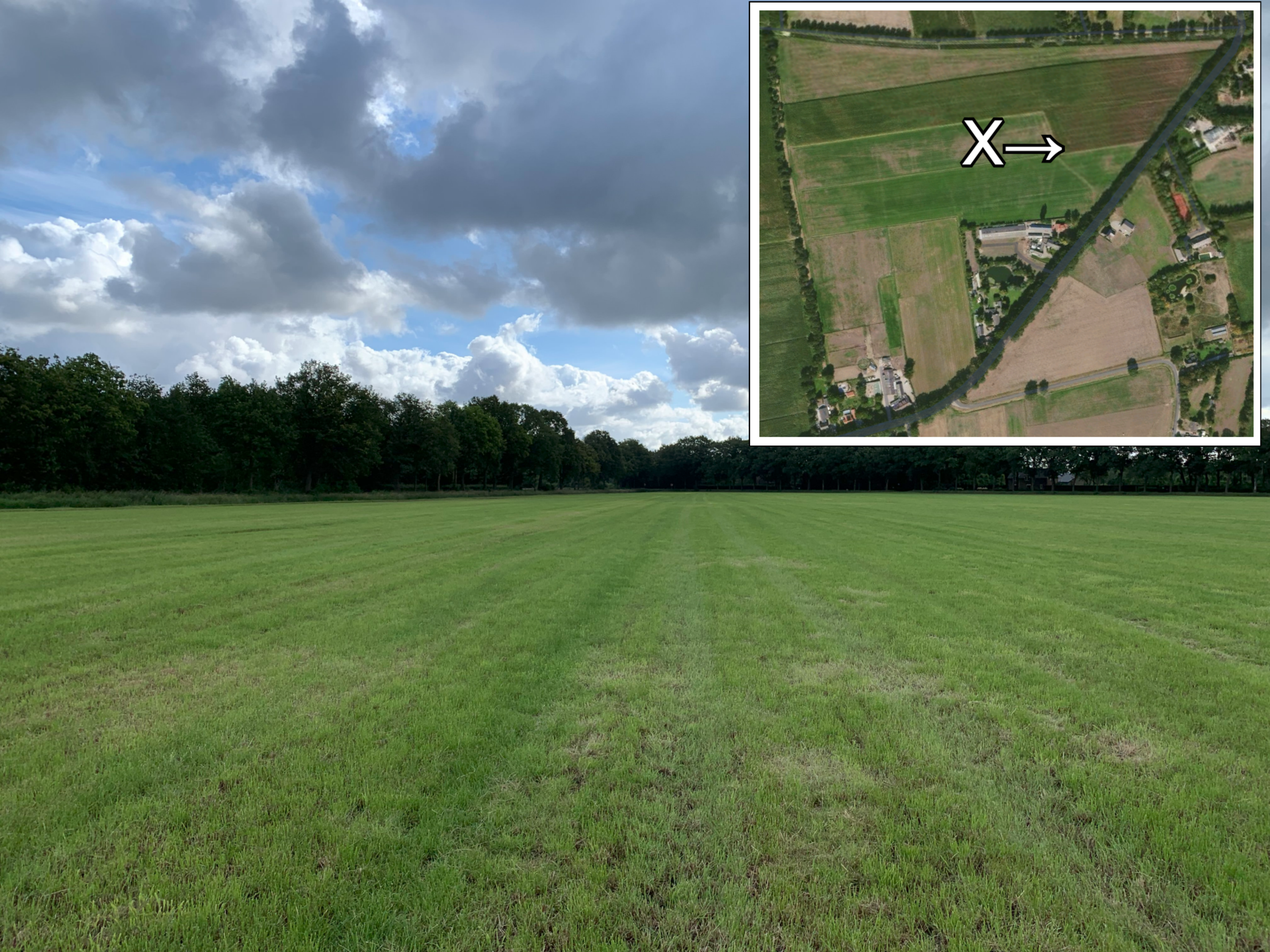}
    \caption{Picture taken at E1 (Outdoor Low-Multipath). The X and arrow on the satellite image (top right) indicate testing location and direction.}
    \label{fig:E1_Visual}
\end{figure}

\textbf{E1: Outdoor Low-Multipath.} This environment is a wild field in the countryside, as shown in Fig. \ref{fig:E1_Visual}, with no obstructions in a radius of $100$~m. The possibility of multipath fading is minimized, although the signal could still reflect from the ground.

\begin{figure}[h]
    \centering
    \includegraphics[width=.8\columnwidth]{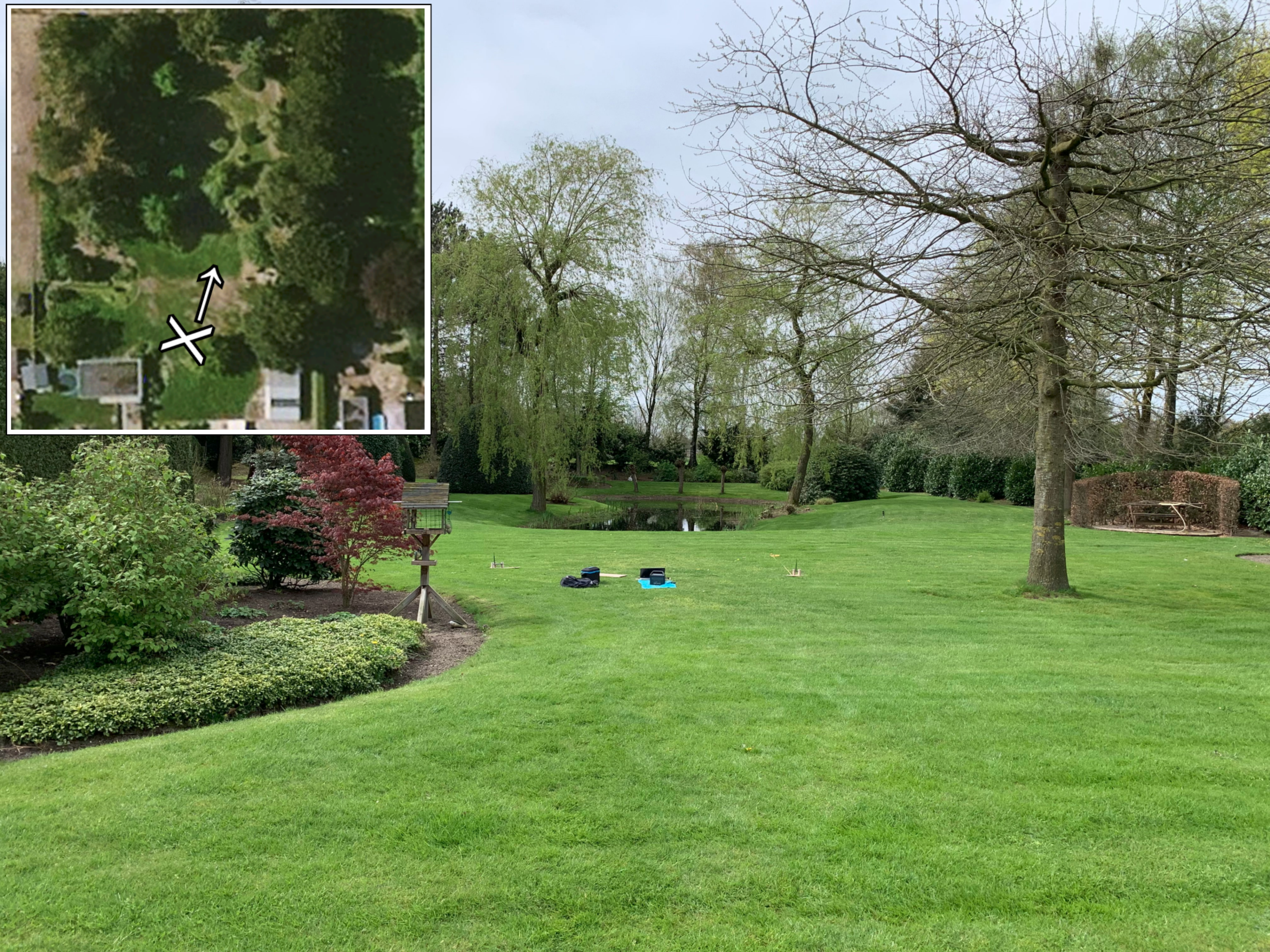}
    \caption{Picture taken at E2 (Outdoor Multipath). The X and arrow on the satellite image (top left) indicate testing location and direction.}
    \label{fig:E2_Visual}
\end{figure}

\textbf{E2: Outdoor Multipath.} This environment is a large garden with trees, plants, and a shed, as shown in Fig. \ref{fig:E2_Visual}. Here, multipath is likely to occur due to the obstacles located very close to the deployment location of the devices.

\begin{figure}[h]
    \centering
    \includegraphics[width=.8\columnwidth]{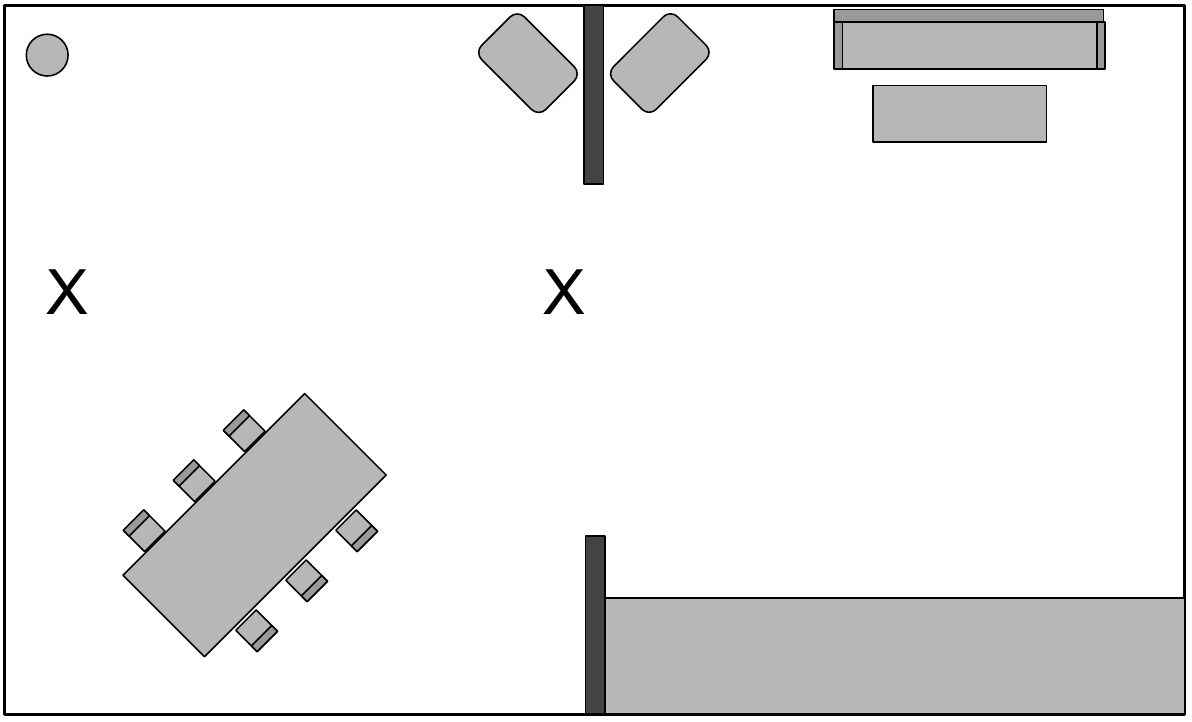}
    \caption{Layout map for E3 (Indoor). The symbols \emph{X} indicate testing locations. The room contains a coathanger (top left), two wood stoves (top center), a table with chairs (bottom left), two walls (top and bottom center), a couch with a table (top right), and a kitchen (bottom right).}
    \label{fig:E3_Visual}
\end{figure}

\begin{figure}[h]
    \centering
    \includegraphics[width=\columnwidth]{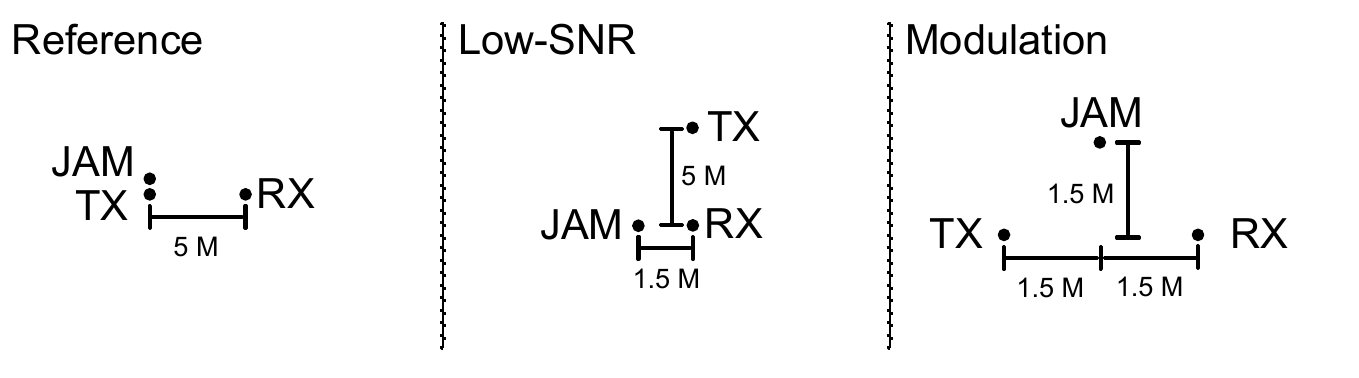}
    \caption{Scenario considered in our measurement campaign: Reference setup (left), Low-SNR (middle), and Modulation (right) physical setup.}
    \label{fig:Ref_SNR_Mod}
\end{figure}

\textbf{E3: Indoor.} This is an office-like environment with a table, chairs, and windows, depicted in Fig. \ref{fig:E3_Visual}. Due to the presence of walls, the RF signal is reflected and attenuated.
We schematize the reference setup of the devices in the experiments in Fig.~\ref{fig:Ref_SNR_Mod}. 

The relative position of the transmitter, receiver, and jammer changes as a function of the scenario considered due to our requirement for weak jamming of the communication link. In the reference scenario, the jammer is in the same position as the transmitter, while both are 5 meters from the receiver. In the Low-SNR scenario, the distance between the transmitter and the receiver is still 5 meters, but the jammer is deployed closer to the receiver (1.5 meters). Finally, in the Modulation scenario, all the devices are 1.5 meters away from a reference point. We considered two channel conditions for each setup, namely, without jamming (NO JAM) and with jamming (JAM). By default, we took measurements lasting 15 minutes for the JAM scenario and 30 minutes for the NO JAM scenario. For the reference setup, we set the gains on the TX, RX, and Jammer to $62$, $50$, and $44$, respectively. These values allow for a stable and reliable communication link, a fundamental condition for our tests. 

\textcolor{black}{We highlight that reproducing \emph{weak jamming} conditions in real-world scenarios is challenging. Indeed, each environment (indoors, outdoors w/o multipath) is characterized by peculiar propagation conditions, which affect the impact of the jammer on the legitimate communication link. Moreover, our measurements have been executed in regular operational conditions, with people and vehicles moving around the communication link and other networks coexisting on the same spectrum. Therefore, it is not possible to control precisely the communication channel--this would also be unfair, since we consider real-world communication environments.
Therefore, all our experiments consider as a \emph{base condition} a situation where the BER of the legitimate communication link under jamming (JAM) is not more than 1\% higher than the BER of the legitimate communication link without jamming (NO JAM). Such a condition, achieved by regulating the jammer gain, allows us to experience the \emph{weak jamming} scenario and investigate what are the capabilities to detect jamming of our proposed solutions and the techniques in the current literature.}

Starting from the reference setup, for our tests, we investigate the following factors: (i) image generation parameters, (ii) jammer parameters, and (iii) communication link parameters. All data collected as part of this work are available at~\cite{data}, while we provide relevant results of our experiments in Sec.~\ref{sec:results}.

\section{Experimental Results}
\label{sec:results}

This section shows the results of our experiments. We first introduce the main performance metrics in Sec.~\ref{sec:metrics}, and then present the results about the impact on jamming detection of various image generation parameters (Sec.~\ref{sec:img_gen_params}), jammer parameters (Sec.~\ref{sec:Res_JOR_HW_TYPE_LOC}), and communication parameters (Sec.~\ref{sec:comm_params}). \textcolor{black}{Finally, we evaluate the overhead of our solution in Sec.~\ref{sec:overhead}. We summarize the details of the experiments in Tab.~\ref{tab:experiments}. Note that we do not aim to compare environments across them, but to show that detecting weak jamming signals is possible in all environments across various configuration and communication parameters.
}
\begin{table*}[!t]
\scriptsize
\centering
\caption{\textcolor{black}{Details of the experiments}.}
\label{tab:experiments}
\color{black}
\begin{tabular}{c|l|l|l|l|l|l}
\multicolumn{1}{l|}{}                                                            & {\bf Exp.} & {\bf Control Variable}                                                    & {\bf Envs.}      & {\bf Values}                                                                                                               & {\bf Other parameters}                                                                                                          & {\bf Approaches}                                                                                                                  \\ \hline
\multirow{3}{*}{\begin{tabular}[c]{@{}c@{}} { \bf Image Gen. }\\  {\bf Parameters}\end{tabular}} & 1    & \begin{tabular}[c]{@{}l@{}}No. of samples \\ per image\end{tabular} & E2         & $10^4$,   $5 \cdot 10^4$, $1 \cdot 10^5$                                                                                                      & \begin{tabular}[c]{@{}l@{}}Jammer Gain=32;  \\ TX Gain= 62; \\ RX Gain= 50\end{tabular}                                   & \begin{tabular}[c]{@{}l@{}}Alhazbi et al. {[}3{]}, \\ Sciancalepore et al. {[}4{]}, \\ Ours - CNN, \\ Ours -AE\end{tabular} \\ \cline{2-7} 
                                                                                  & 2    & \begin{tabular}[c]{@{}l@{}}Training \\ Set Size\end{tabular}        & E2            & 2,9,18,36,54,72,120                                                                                                  & \begin{tabular}[c]{@{}l@{}}Samples per image=$1 \cdot 10^5$; \\ Jammer Gain=44;  \\ TX Gain= 62; \\ RX Gain= 50\end{tabular}         & \begin{tabular}[c]{@{}l@{}}Alhazbi et al. {[}3{]}, \\ Sciancalepore et al. {[}4{]}, \\ Ours - CNN, \\ Ours -AE\end{tabular} \\ \cline{2-7} 
                                                                                  & 3    & \begin{tabular}[c]{@{}l@{}}Augment. \\ Strategy\end{tabular}        & E2         & \begin{tabular}[c]{@{}l@{}}180-deg. rot., \\ left-right mirror, \\ up-down mirror, \\ contrast, \\ brightness\end{tabular} & \begin{tabular}[c]{@{}l@{}}Samples per image=$5 \cdot 10^4$; \\ Jammer Gain=44;  \\ TX Gain= 62; \\ RX Gain= 50\end{tabular}         & \begin{tabular}[c]{@{}l@{}}Alhazbi et al. {[}3{]}, \\ Sciancalepore et al. {[}4{]}, \\ Ours - CNN, \\ Ours -AE\end{tabular} \\ \hline
\multirow{5}{*}{\begin{tabular}[c]{@{}c@{}} {\bf Jamming} \\ {\bf Parameters}\end{tabular}}    & 4    & \begin{tabular}[c]{@{}l@{}}Jammer\\ Oversamp.\\ Ratio\end{tabular}  & E1, E2, E3 & 1,   2, 3, 4                                                                                                         & \begin{tabular}[c]{@{}l@{}}Samples \\ per image=$1 \cdot 10^5$; \\ Jammer Gain=44;  \\ TX Gain= 62; \\ RX Gain= 50\end{tabular}      & \begin{tabular}[c]{@{}l@{}}Ours - CNN, \\ Ours   -AE\end{tabular}                                                           \\ \cline{2-7} 
                                                                                  & 5    & \begin{tabular}[c]{@{}l@{}}Jammer  \\ Hardware\end{tabular}         & E2, E3     & \begin{tabular}[c]{@{}l@{}}LimeSDR   Mini \\ LimeSDR USB  \end{tabular}                                                                                        & \begin{tabular}[c]{@{}l@{}}Samples per image=$1 \cdot 10^5$;\\ Jammer Gain=44; \\ TX Gain= 62; \\ RX Gain= 50\end{tabular}           & \begin{tabular}[c]{@{}l@{}}Ours - CNN, \\ Ours   -AE\end{tabular}                                                           \\ \cline{2-7} 
                                                                                  & 6    & \begin{tabular}[c]{@{}l@{}}Jammer \\ Signal \\ Type\end{tabular}    & E1, E2, E3 & AWGN,   Deceptive                                                                                                    & \begin{tabular}[c]{@{}l@{}}Samples per image=$1 \cdot 10^5$; \\ Jammer Gain=44;  \\ TX Gain= 62;\\ RX Gain= 50\end{tabular}          & \begin{tabular}[c]{@{}l@{}}Ours - CNN, \\ Ours   -AE\end{tabular}                                                           \\ \cline{2-7} 
                                                                                  & 7    & \begin{tabular}[c]{@{}l@{}}Jammer \\ Location\end{tabular}          & E1, E2, E3 & 1,   2, 3,                                                                                                           & \begin{tabular}[c]{@{}l@{}}Samples per image=$1 \cdot 10^5$; \\ Jammer Gain=44;  \\ TX Gain= 62; \\ RX Gain= 50\end{tabular}         & \begin{tabular}[c]{@{}l@{}}Ours - CNN, \\ Ours   -AE\end{tabular}                                                           \\ \cline{2-7} 
                                                                                  & 8    & \begin{tabular}[c]{@{}l@{}}Jammer  \\ Distance\end{tabular}         & E1, E2     & \begin{tabular}[c]{@{}l@{}} 3,   5, 7, 10, 13, 16,\\ 19, 21, 23, 25  \end{tabular}                                                                              & \begin{tabular}[c]{@{}l@{}}Samples per image=$1 \cdot 10^5$; \\ Jammer Gain=44;  \\ TX Gain= 62; \\ RX Gain= 50\end{tabular}         & \begin{tabular}[c]{@{}l@{}}Alhazbi et al. {[}3{]}, \\ Sciancalepore et al. {[}4{]}, \\ Ours - CNN, \\ Ours -AE\end{tabular} \\ \hline
\multirow{4}{*}{\begin{tabular}[c]{@{}c@{}} { \bf Commun.} \\ {\bf Parameters}\end{tabular}}    & 9    & \begin{tabular}[c]{@{}l@{}}Mod. \\ Scheme\end{tabular}              & E2         & \begin{tabular}[c]{@{}l@{}}BPSK,  QPSK, \\ 16-QAM, \\ 64-QAM\end{tabular}  & \begin{tabular}[c]{@{}l@{}}Samples per image=$1 \cdot 10^5$; \\ Jammer Gain=66,50,16;    \\ TX Gain= 62; \\ RX Gain= 50\end{tabular} & \begin{tabular}[c]{@{}l@{}}Alhazbi et al. {[}3{]}, \\ Sciancalepore et al. {[}4{]}, \\ Ours - CNN, \\ Ours -AE\end{tabular} \\ \cline{2-7} 
                                                                                  & 10   & \begin{tabular}[c]{@{}l@{}}Trans. \\ Gain\end{tabular}              & E2, E3     & \begin{tabular}[c]{@{}l@{}} {[}50,   55, 60, 65{]}, \\ {[}44, 49, 54, 59{]}  \end{tabular}                                                                       & \begin{tabular}[c]{@{}l@{}}Samples per image=$1 \cdot 10^5$;  \\ Jammer Gain=44; \\ TX Gain= 14; \\ RX Gain=50;\end{tabular}         & \begin{tabular}[c]{@{}l@{}}Alhazbi et al. {[}3{]}, \\ Sciancalepore et al. {[}4{]}, \\ Ours - CNN, \\ Ours -AE\end{tabular} \\ \cline{2-7} 
                                                                                  & 11   & \begin{tabular}[c]{@{}l@{}}Recv. \\ Dist.\end{tabular}              & E1, E2     & \begin{tabular}[c]{@{}l@{}} 5,   7, 10, 13, 16, 19, \\ 21, 23, 25, 27 \end{tabular}                                                                               & \begin{tabular}[c]{@{}l@{}}Samples per image=$1 \cdot 10^5$; \\ Jammer Gain=66,50,16;   \\ TX Gain= 62; \\ RX Gain= 50\end{tabular}  & \begin{tabular}[c]{@{}l@{}}Alhazbi et al. {[}3{]}, \\ Sciancalepore et al. {[}4{]}, \\ Ours - CNN, \\ Ours -AE\end{tabular} \\ \cline{2-7} 
                                                                                  & 12   & \begin{tabular}[c]{@{}l@{}}Recv.\\ Mob.\end{tabular}                & E2         & \begin{tabular}[c]{@{}l@{}} Static, \\  Parallel, \\ Perpendicular \end{tabular}                                                                                    & \begin{tabular}[c]{@{}l@{}}Samples per Image=$1 \cdot 10^5$;  \\ Jammer Gain=44; \\ TX Gain=62; \\ RX Gain=50,\end{tabular}          & \begin{tabular}[c]{@{}l@{}}Alhazbi et al. {[}3{]}, \\ Sciancalepore et al. {[}4{]}, \\ Ours - CNN, \\ Ours -AE\end{tabular} \\ 
\end{tabular}
\end{table*}
\subsection{Performance Metrics}
\label{sec:metrics}
We use \emph{accuracy} as the main performance metric, computed as $\frac{TP + TN}{TP + FP + TN + FN}$, being $TP$ the number of true positives (\emph{JAM} images correctly classified), $TN$ the true negatives (\emph{NO JAM} images correctly classified), $FP$ the number of false positives (\emph{NO JAM} images classified as \emph{JAM}) and $FN$ the number of false negatives (\emph{JAM} images classified as \emph{NO JAM}). Although such data are available, we do not report precision and recall results since we are interested in comparing the overall performance of the investigated approaches rather than the contribution of False Positive and False Negative. To obtain our results, we use K-fold cross-validation with $K = 5$ and, for each value, we report the mean and $95$\% confidence intervals, computed using Matlab's $tinv$ function, i.e.,  the inverse cumulative distribution function of the Student's distribution. The confidence intervals are represented as red error bars in the bar plots, with a different color for each result in the line plots. \textcolor{black}{To ensure fairness of comparison between \acp{CNN} and \acp{AE}, we used the same amount of images for training both classifiers, i.e. $150$ \emph{NO JAM} images. Then, for each experiment, the testing set is constituted by an equal number of \emph{NO JAM} and \emph{JAM} images, depending the amount of jamming data available for the particular experiment. This number can be obtained from Tab.~\ref{tab:dataset}, dividing the number of IQ samples for each environment and testing conditions by the number of samples per image used. The remaining part of the \emph{NO JAM} dataset has been used for validation.}

For some plots, we introduce a custom metric, i.e., the \emph{\ac{SNR} Degradation Ratio}, typically presented as the second x-axis. The \emph{\ac{SNR} Degradation Ratio} models the impact of jamming on the communication channel, and it is computed as the ratio between the average \ac{SNR} of the \emph{JAM} images and the average \ac{SNR} of the \emph{NO JAM} images, i.e., $SNR_{DR} = \frac{SNR_{JAM}}{SNR_{NOJAM}}$. Note that, due to different factors affecting SNR, the SNR Degradation ratio is not comparable across different experiments. Finally, whenever not specified, we use only measurements characterized by a \ac{BER} $< 0.01$, so to enforce a \emph{low-BER} scenario following the definition in Sec.~\ref{sec:background}.

\subsection{Image Generation Parameters}
\label{sec:img_gen_params}
In the following subsection, we investigate the impact of various image generation parameters on jamming detection accuracy, including samples per image, training set size, and image augmentation strategies.

\begin{figure}[h]
    \centering
    \includegraphics[width=0.48\textwidth]{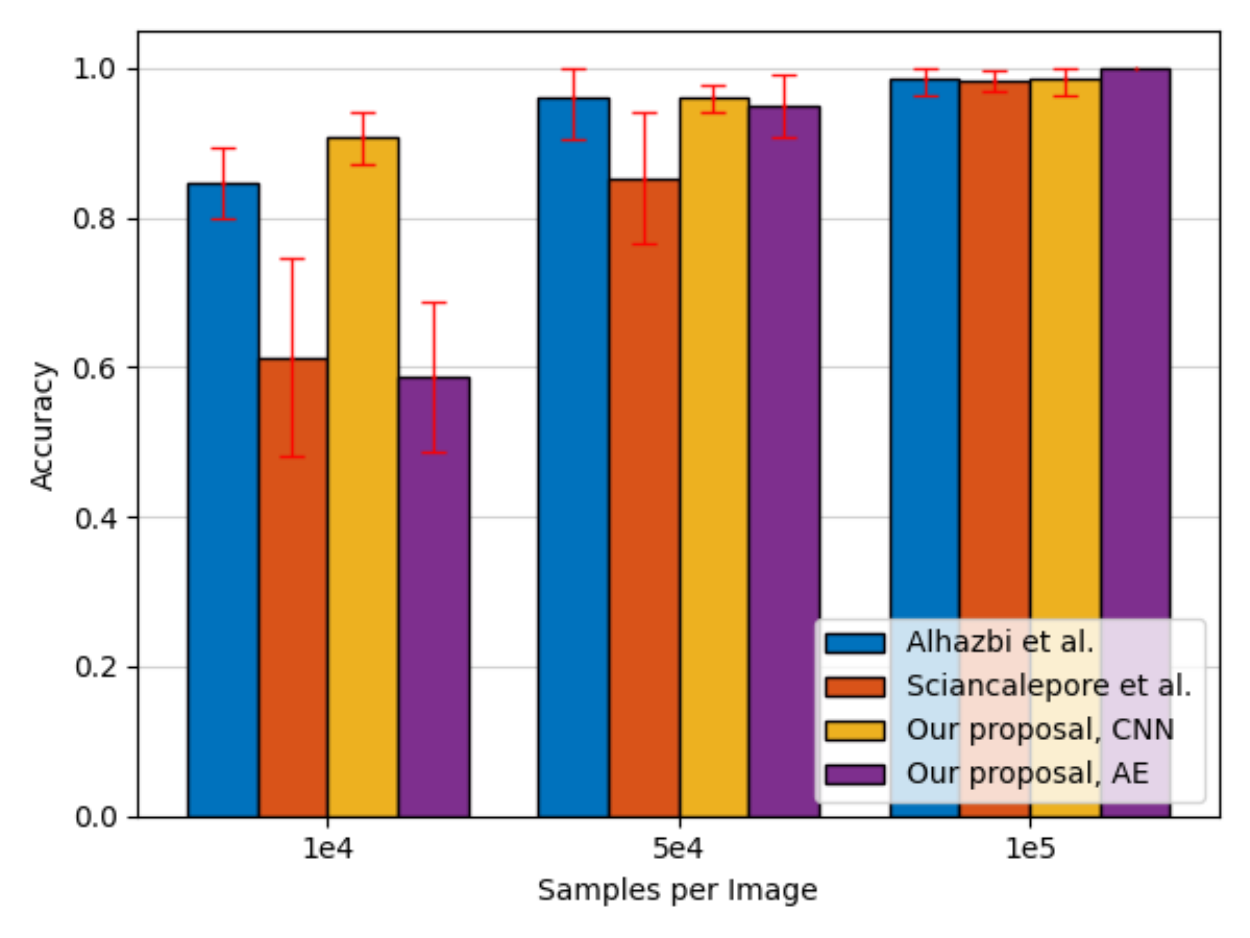}
    \caption{Accuracy as a function of the number of samples per image. We compared our proposed solutions with state-of-the-art contributions.}
    \label{fig:i1_i2_SampPerImg}
\end{figure}

{\bf Exp. 1: Samples per Image.} To provide insight into the effect of the number of samples per image $n$ on the accuracy of jamming detection, we carried out experiments for $n \in [1^{4}, 5^{4}, 1^{5}]$. At any higher values for $n$, the number of images produced is too small to allow meaningful results. Figure~\ref{fig:i1_i2_SampPerImg} shows the accuracy as a function of the number of samples per image while comparing our solutions (\ac{CNN} and AE) with the state of the art~\cite{alhazbi2023_ccnc} and~\cite{sciancalepore2024_iotj}, considering our reference setup for environment $E2$ and a jammer gain value of $32$. 


\textcolor{black}{We notice that the higher the number of samples per image, the higher the accuracy for all approaches. Specifically, the \ac{CNN}-based approach improves from an accuracy of 0.9 with $10^4$ samples per images to an average accuracy of 0.98 with $10^5$ samples. Similarly, the \ac{AE}-based approach improves from 0.58 with $10^4$ samples to 0.995 with $10^5$ samples.}
We notice that the \ac{CNN}-based approach requires fewer samples to reach very high accuracy (higher than $0.95$), i.e., $5\cdot10^4$ samples per image. In such conditions, we also notice that our proposed approach significantly outperforms both previous solutions. We adopted $10^5$ samples per image for the rest of our analysis. \textcolor{black}{Consider that we collected data with a sample rate of $5\cdot 10^6$~\ac{sps}. Thus, the time necessary to collect $10^5$ samples to generate an image is $\frac{1 \cdot 10^5}{5 \cdot 10^6} = 0.02$~s, making our solution very suitable for real-time jamming detection.}

\begin{figure}[h]
    \centering
    \includegraphics[width=0.48\textwidth]{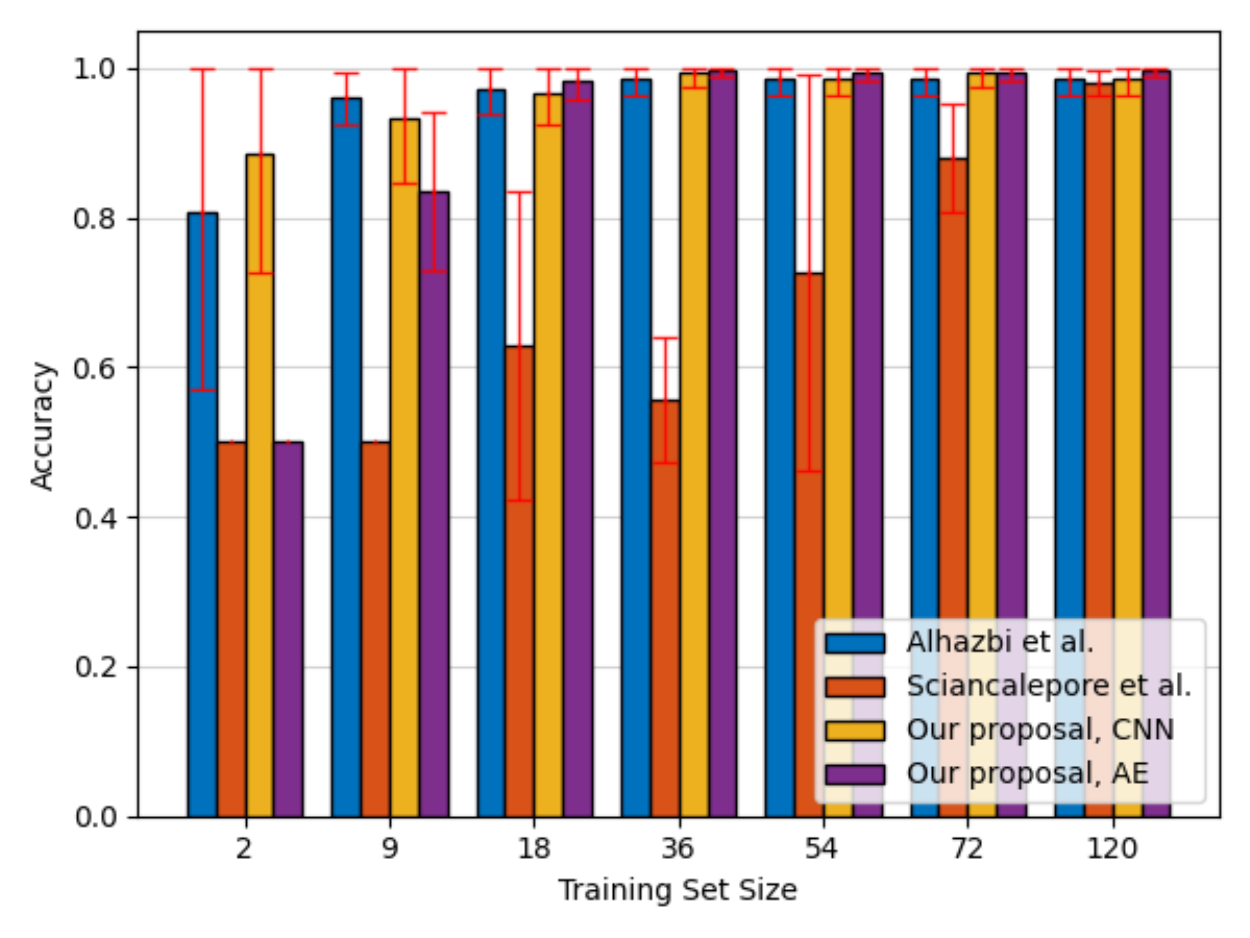}
    \caption{Accuracy as a function of the training set size. We compared our solutions with state-of-the-art contributions.}
    \label{fig:i1_i2_TrainSize_CNN_AE}
\end{figure}

{\bf Exp. 2: Training Set Size.} We also investigate the impact of the training set size on the classification accuracy. The training strategy of the two approaches is different. When the \ac{AE}-based anomaly detection approach uses $s$ images for training, such images are all images built from non-jammed samples; instead, when the CNN-based binary classification approach uses $s$ images, we use $\frac{s}{2}$ images for both jamming and no-jamming. We report the results of our investigation in Fig.~\ref{fig:i1_i2_TrainSize_CNN_AE}. 

Jamming detection accuracy is positively correlated with the number of training images. However, using the approach in~\cite{sciancalepore2024_iotj} requires more images during the training phase to reach stable performance, i.e., $120$ images, compared to the only $18$ images required by our proposal. Jamming detection performances with our approach are very reliable when using $18$ images, with both \acp{CNN} and \acp{AE}.

\begin{figure}
\centering
    \subfloat[Sciancalepore et al.,~\cite{sciancalepore2024_iotj}\label{fig:i1_ImgAug_5e4_AE}]{%
    \includegraphics[width=.8\columnwidth]{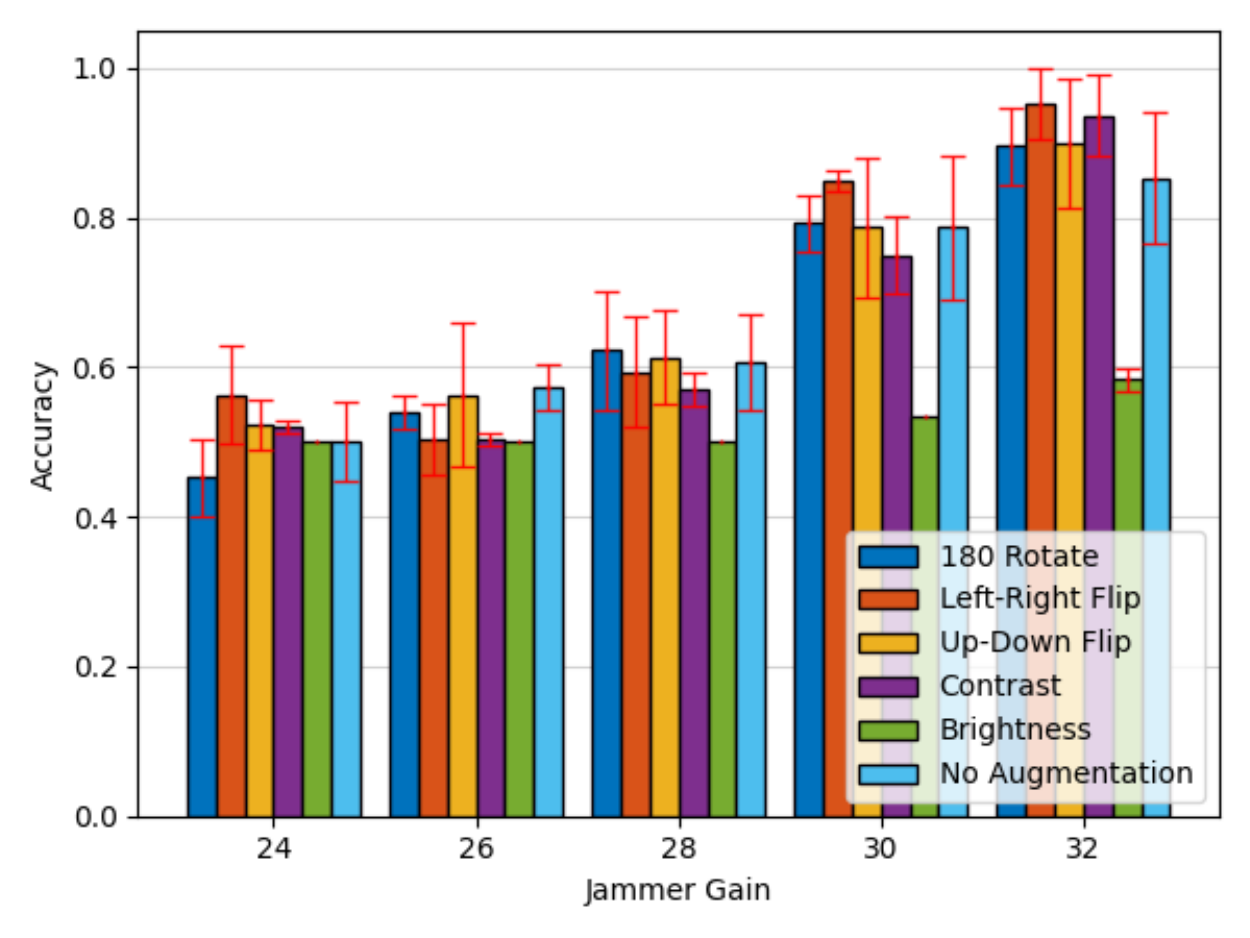}}
    \hfill
    \subfloat[Alhazbi et al.,~\cite{alhazbi2023_ccnc}\label{fig:i1_ImgAug_5e4_CNN}]{%
    \includegraphics[width=.8\columnwidth]{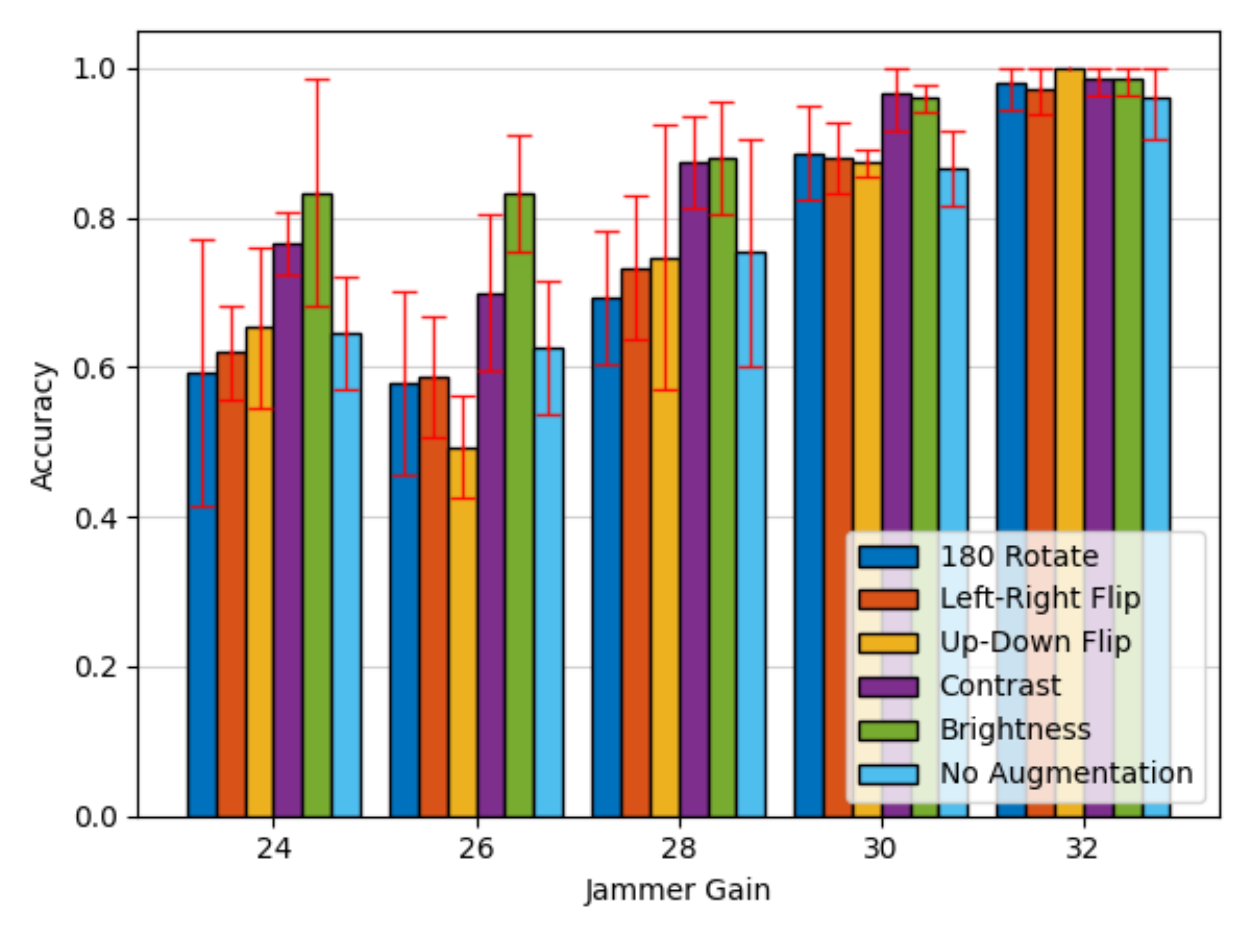}}
    \hfill
    \subfloat[Our proposal, \acp{AE}\label{fig:i2_ImgAug_5e4_AE}]{
    \includegraphics[width=.8\columnwidth]{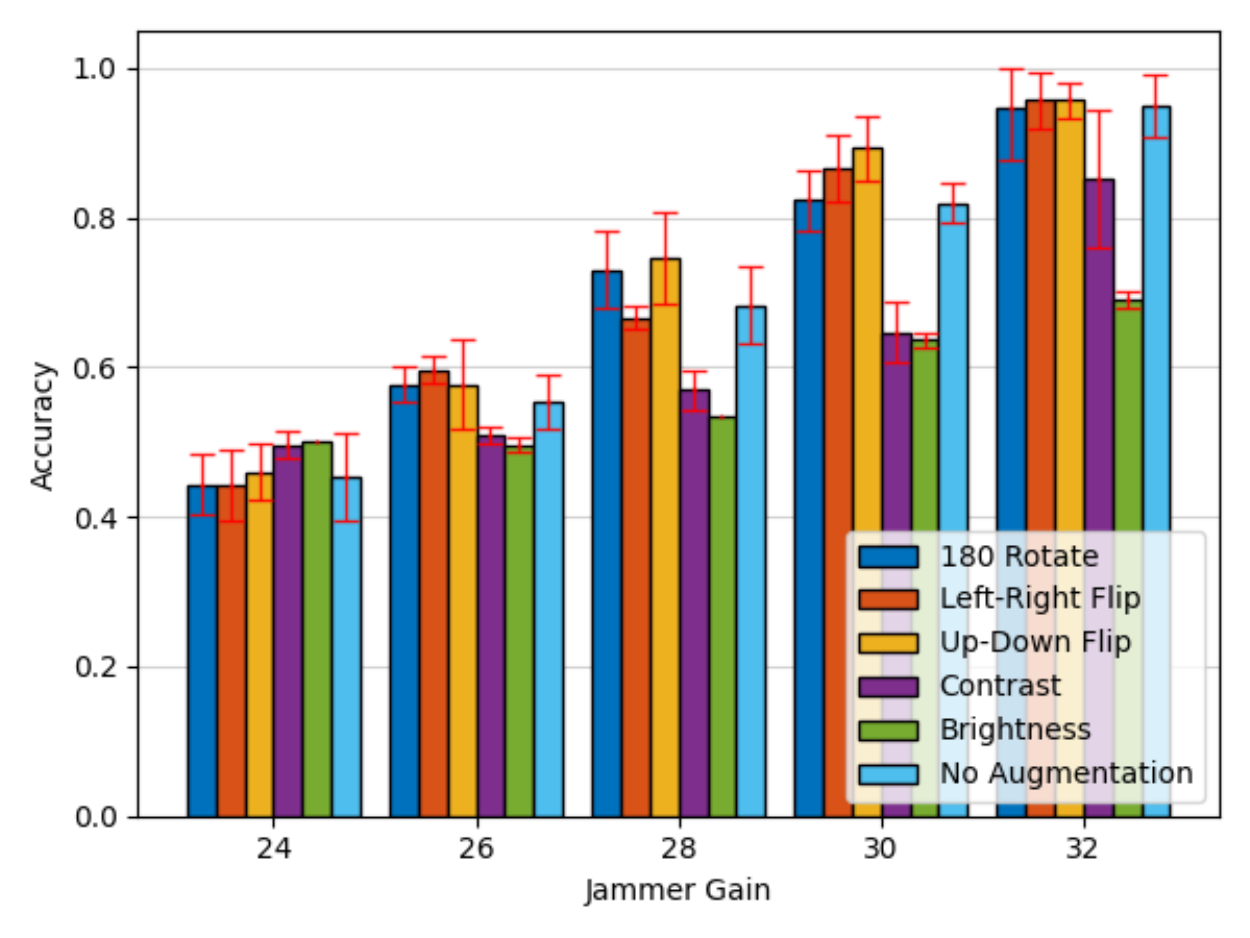}}
    \hfill
    \subfloat[Our proposal, \ac{CNN}\label{fig:i2_ImgAug_5e4_CNN}]{%
    \includegraphics[width=.8\columnwidth]{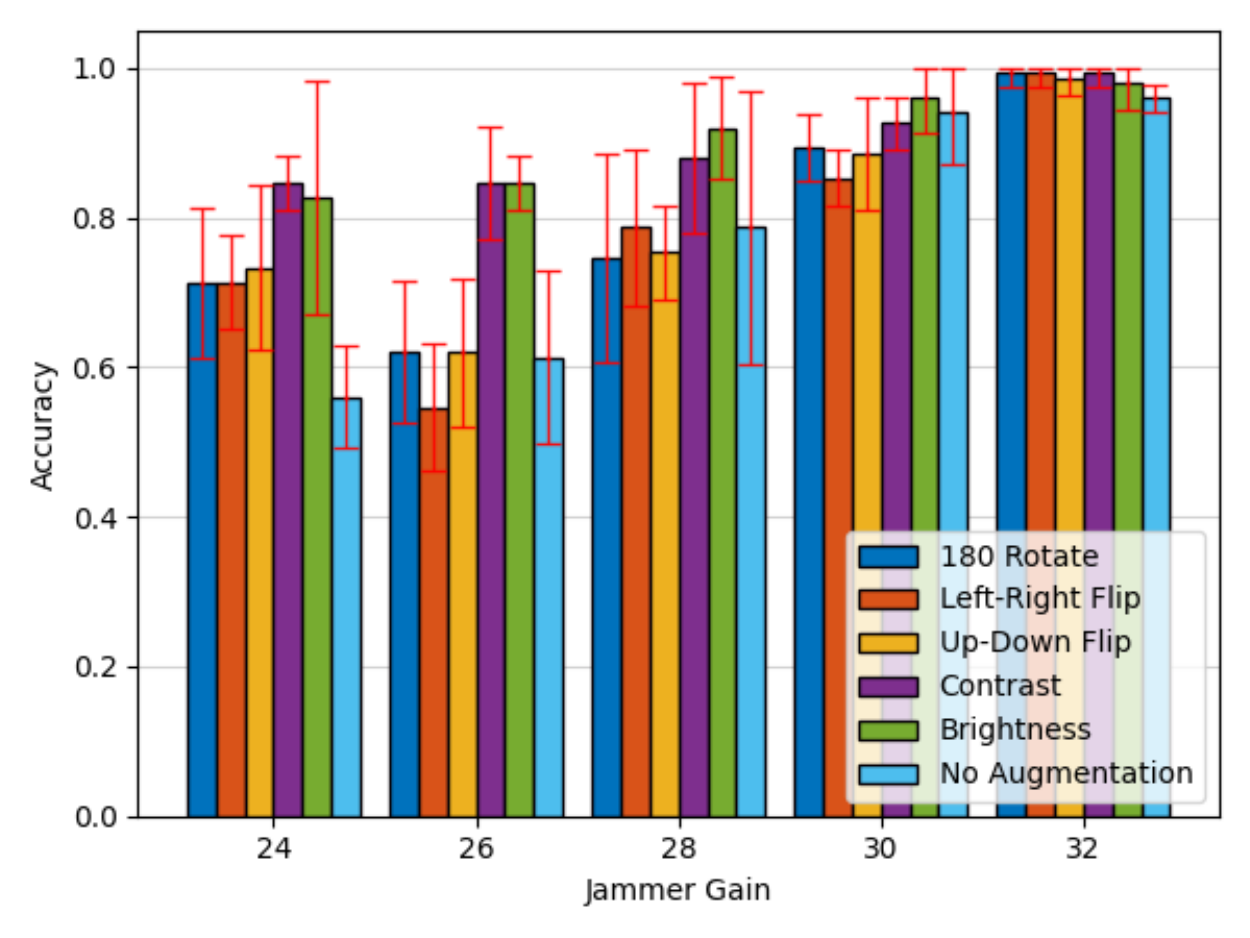}}
    \caption{Accuracy using various image augmentation techniques while considering $5 \cdot 10^4$ samples per image.}
    \label{fig:aug}
\end{figure}

{\bf Exp. 3: Image Augmentation.} We compare the accuracy of our proposed solutions with the literature, considering the adoption of various image augmentation techniques. For this test, we consider the data collected in the environment \emph{E2}, as it provides a more challenging classification task, allowing us to test our solution in the worst-case scenario. We empirically select 150 images from the no-jamming dataset and 75 from the jamming dataset and investigate the effect of image augmentation against the benchmark proposals in~\cite{alhazbi2023_ccnc} and~\cite{sciancalepore2024_iotj} with $5 \cdot 10^4$ samples per image. 
We consider the following augmentation strategies, empirically chosen from those made available by Matlab~\cite{matlab_augmentation}: (i) 180 degrees rotation, (ii) left-right mirror, (iii) up-down mirror, (iv) contrast increase and (v) brightness increase. We report all our results in Fig.~\ref{fig:aug}. We report with the legend \emph{No Augmentation} the performances obtained when only real-world data are used in the training set, providing a reference benchmark. 

\textcolor{black}{When using the \ac{AE}-based one-class classification approach, as in Fig~\ref{fig:aug}(a) and (c), the augmentation strategies \emph{180 Rotate} (dark blue bar), \emph{Left-Right Flip} (orange bar) and \emph{Up-Down Flip} (yellow bar) perform similarly to each other and to using no augmentation, e.g., reporting an accuracy of 0.78 with a jammer gain value of 30 in Fig.~\ref{fig:aug}(a). This indicates that the usage of such augmentation strategies does not decrease performances. The augmentation strategy \emph{Brightness} (green bar) performs significantly worse, e.g., accuracy 0.5 with jammer gain value 30 in Fig.~\ref{fig:aug}(a). \emph{Contrast} works better for the benchmark approaches than ours, e.g., accuracy of 0.96 with jammer gain value 30 in Fig.~\ref{fig:aug}(a).
\emph{Contrast} and \emph{Brightness} perform the best for the CNN-based binary classification approach: \ac{CNN} are exposed to both classes during training. The brightness of the image is correlated with the classification accuracy: bright clouds resemble no jamming, and dim clouds look like jamming. Therefore, using brightness as an augmentation parameter decreases the performance.
Therefore, when choosing the right type of strategy, using augmentation during training does not decrease performance compared to using only real-world data (no augmentation). Thus, augmentation can be used as an effective strategy to increase the dataset when insufficient input data is available. }

\subsection{Jamming Parameters}
\label{sec:Res_JOR_HW_TYPE_LOC}
In this subsection, we investigate the impact of various jamming parameters on the detection of weak jamming signals in the outdoor environment, including the jammer oversampling ratio, the jammer type, the jammer hardware, the jammer location, and the jammer distance from the communication link.

\begin{figure}
    \centering
    \includegraphics[width=0.48\textwidth]{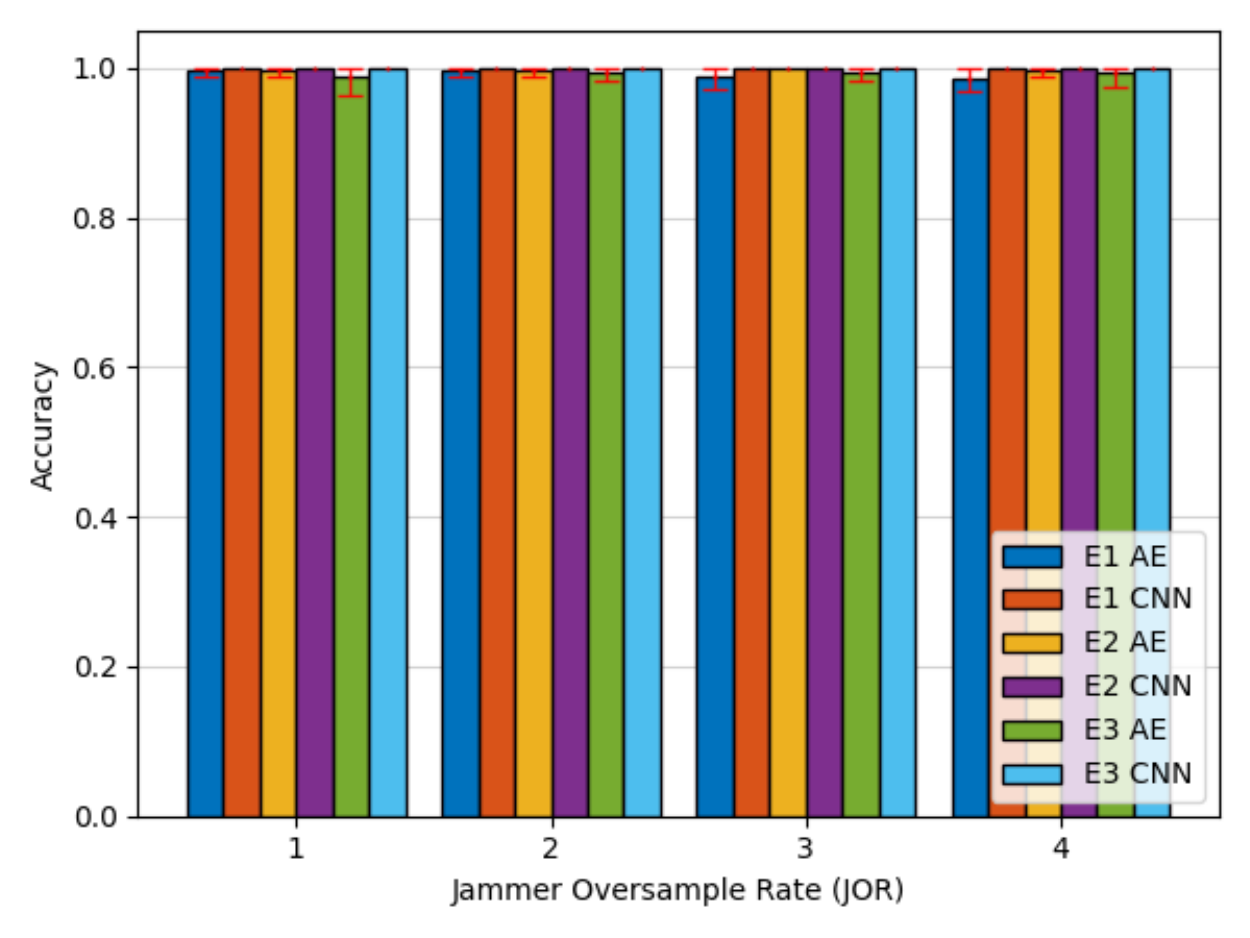}
    \caption{Classification accuracy varying the Jamming Oversampling Ratio (JOR) in environments \emph{E1}, \emph{E2} and \emph{E3}. JOR does not affect the performance of jamming detection.}
    \label{fig:Jam_Samp_Res_i2}
\end{figure}

{\bf Exp. 4: Jammer Oversampling Ratio.} We define \emph{Jammer Oversampling Ratio} (JOR) as the ratio between the sample rate of the jammer and the sample rate of the legitimate communication link. Figure~\ref{fig:Jam_Samp_Res_i2} reports the results of our analysis considering both the proposed jamming detection approaches in the three measurement scenarios.
The accuracy is near $1.0$ for all environments and both detection approaches, demonstrating that increasing the sample rate is not a viable option for the jammer to avoid detection.

\begin{figure}[h]
\centering
    \subfloat[Jammer hardware]{\includegraphics[width=.8\columnwidth]{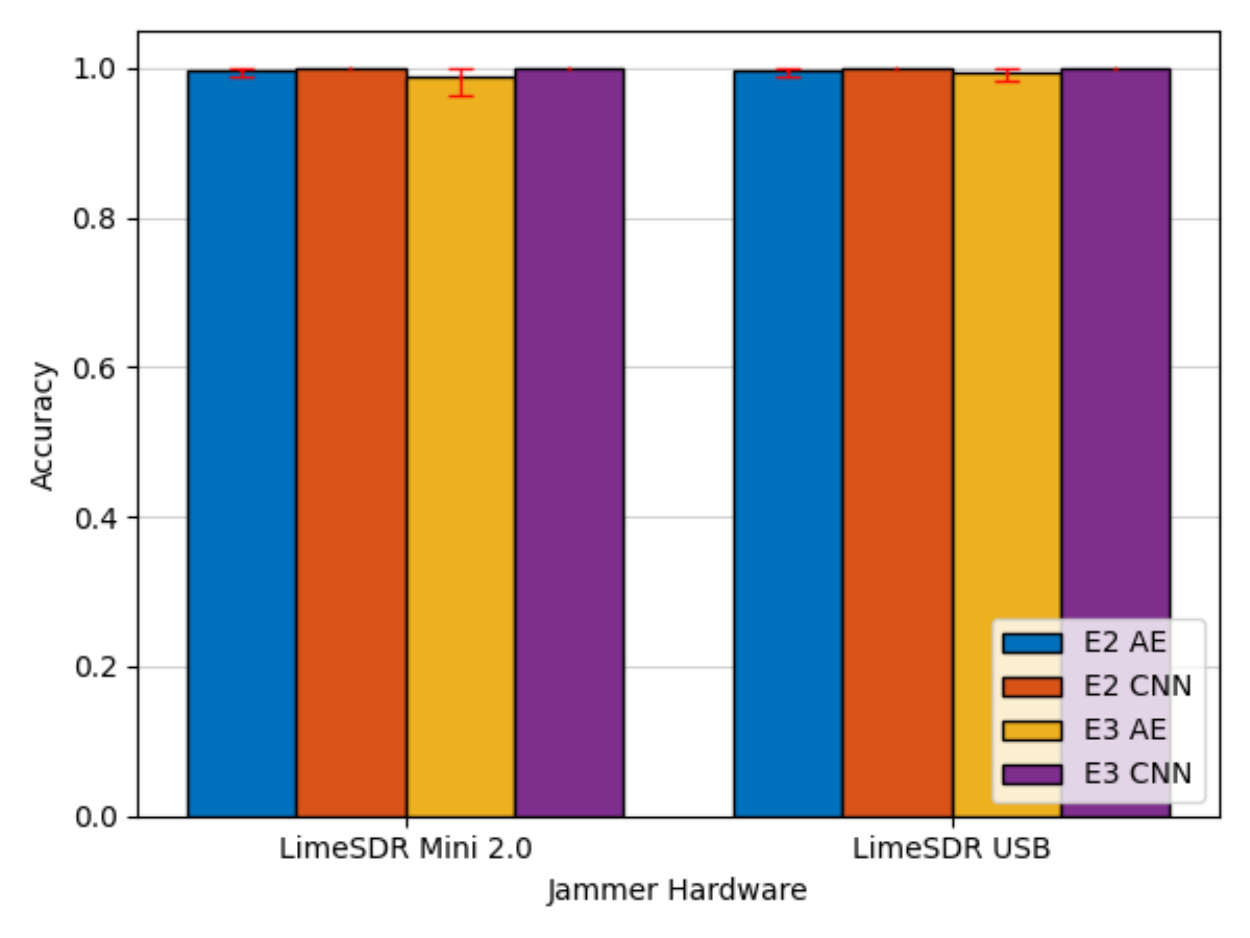}}
    \hfill
    \subfloat[Jamming signal]{\includegraphics[width=.8\columnwidth]{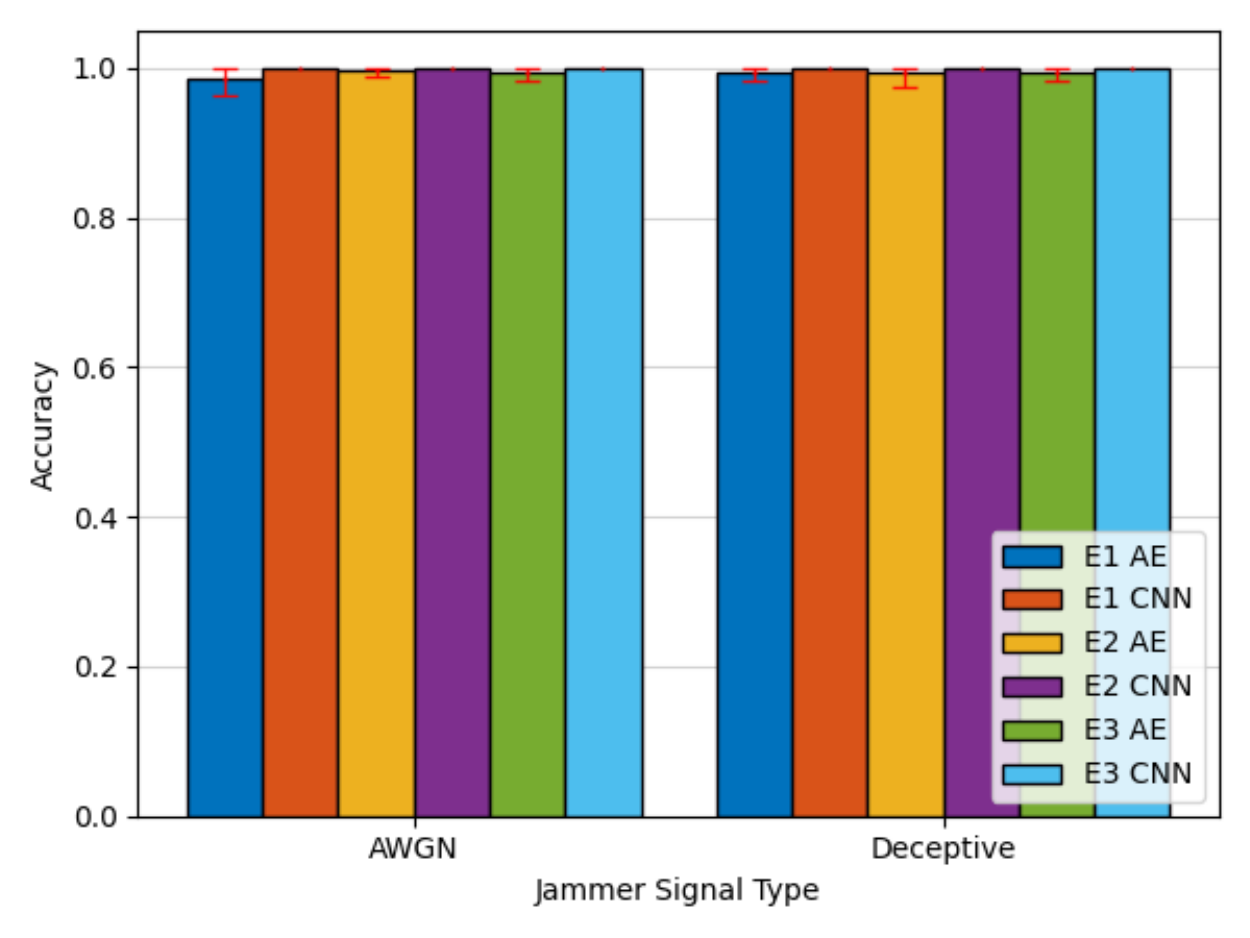}}
    \caption{Performance of jamming detection as a function of the (jammer) hardware and (jamming) signal type.}
    \label{fig:jam_hw_sig}
\end{figure}

{\bf Exps. 5-7: Jammer Hardware, Location, and Type.} We report in Figs.~\ref{fig:jam_hw_sig}(a) and (b) the results of experiments investigating the impact of the \emph{Jammer Hardware} and the \emph{Jammer Signal Type}. 
Our results show that the considered hardware (LimeSDR Mini or LimeSDR) and the different jamming signals (AWGN or deceptive) do not affect the accuracy of our proposed techniques.

\begin{figure}[h]
\centering
    \subfloat[Measurement setup]{\includegraphics[width=\columnwidth]{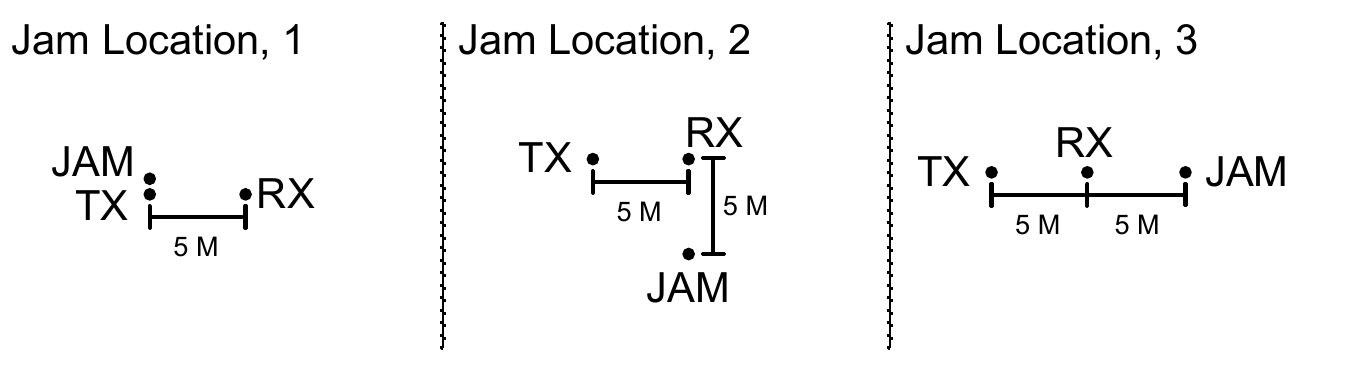}\newline\newline}
    \hfill
    \subfloat[Detection performance]{\includegraphics[width=\columnwidth]{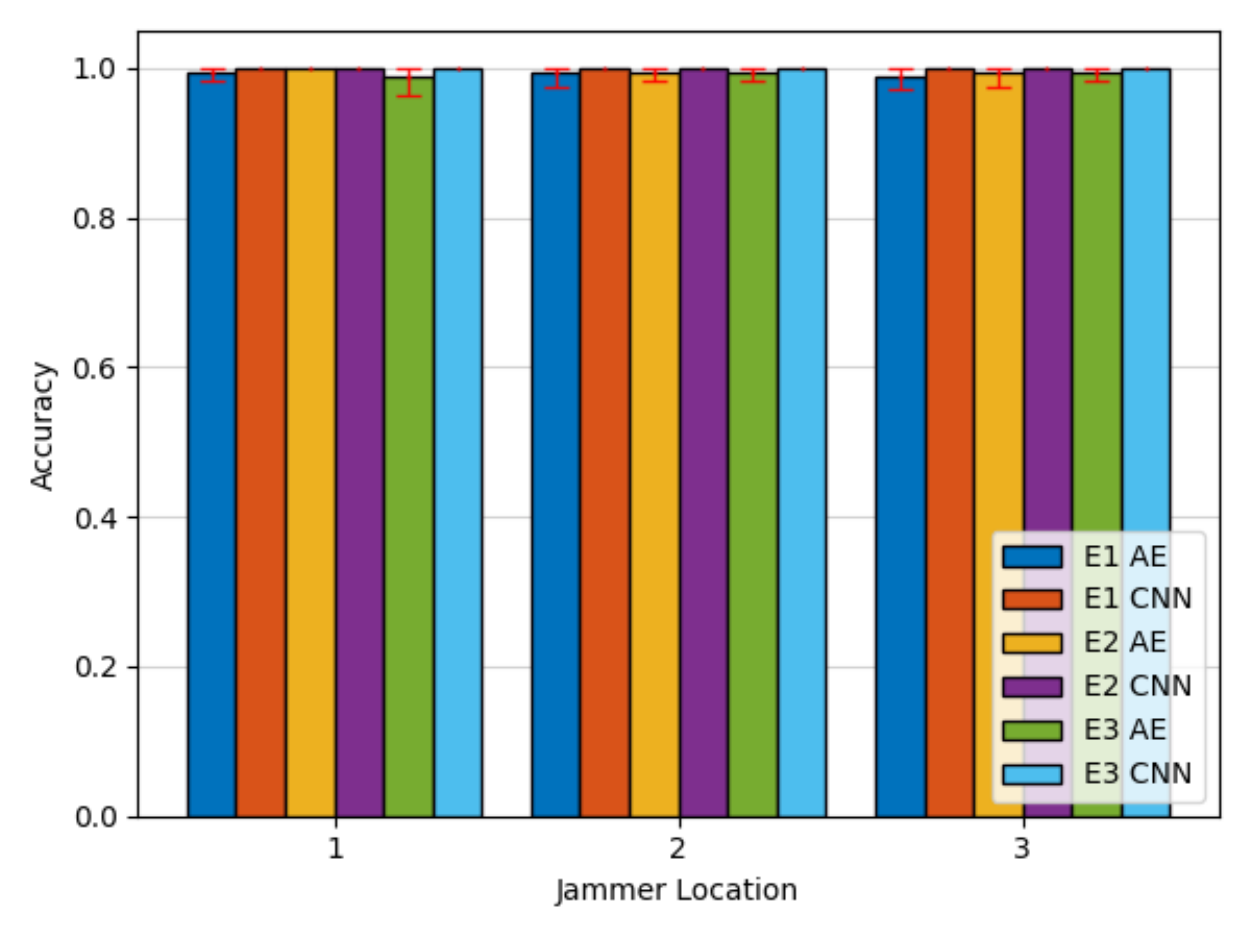}}
    \caption{Detection performance as a function of the relative positions of the jammer and the transmitter-receiver link.}
    \label{fig:jam_loc}
\end{figure}

For the experiments on jammer location, we report in Fig.~\ref{fig:jam_loc}(a) the schematic of the layout used for the experiments. Figure~\ref{fig:jam_loc}(b) shows the results associated with the detection accuracy, confirming the robustness of our proposed solution in any relevant real-world environment. 

\begin{figure}[h]
    \centering
    \includegraphics[width=\columnwidth]{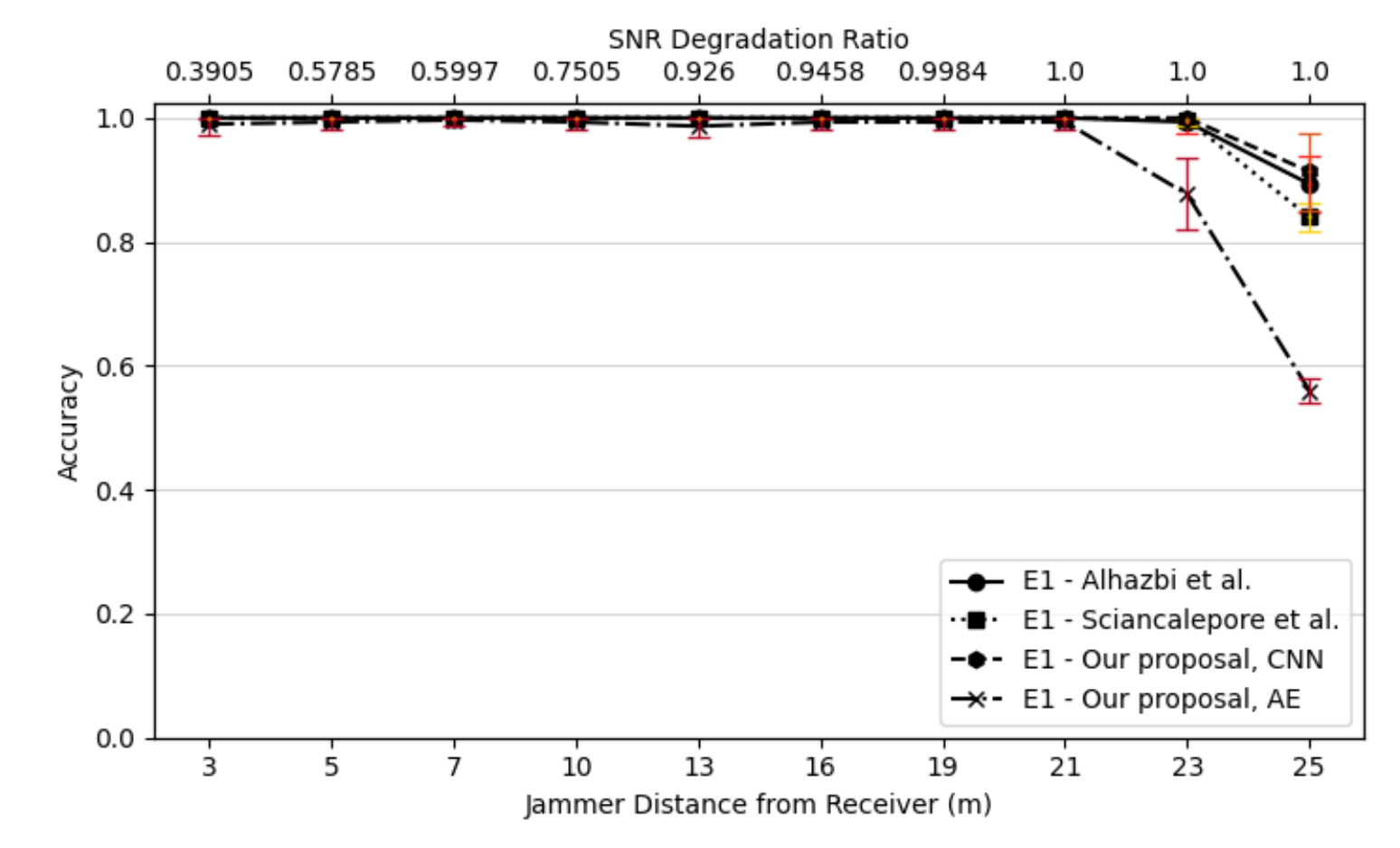}
    \caption{Classification accuracy varying the \emph{Jammer Distance} in environment \emph{E1}.}
    \label{fig:E1_JDist_Res}
\end{figure}

\begin{figure}[h]
    \centering
    \includegraphics[width=\columnwidth]{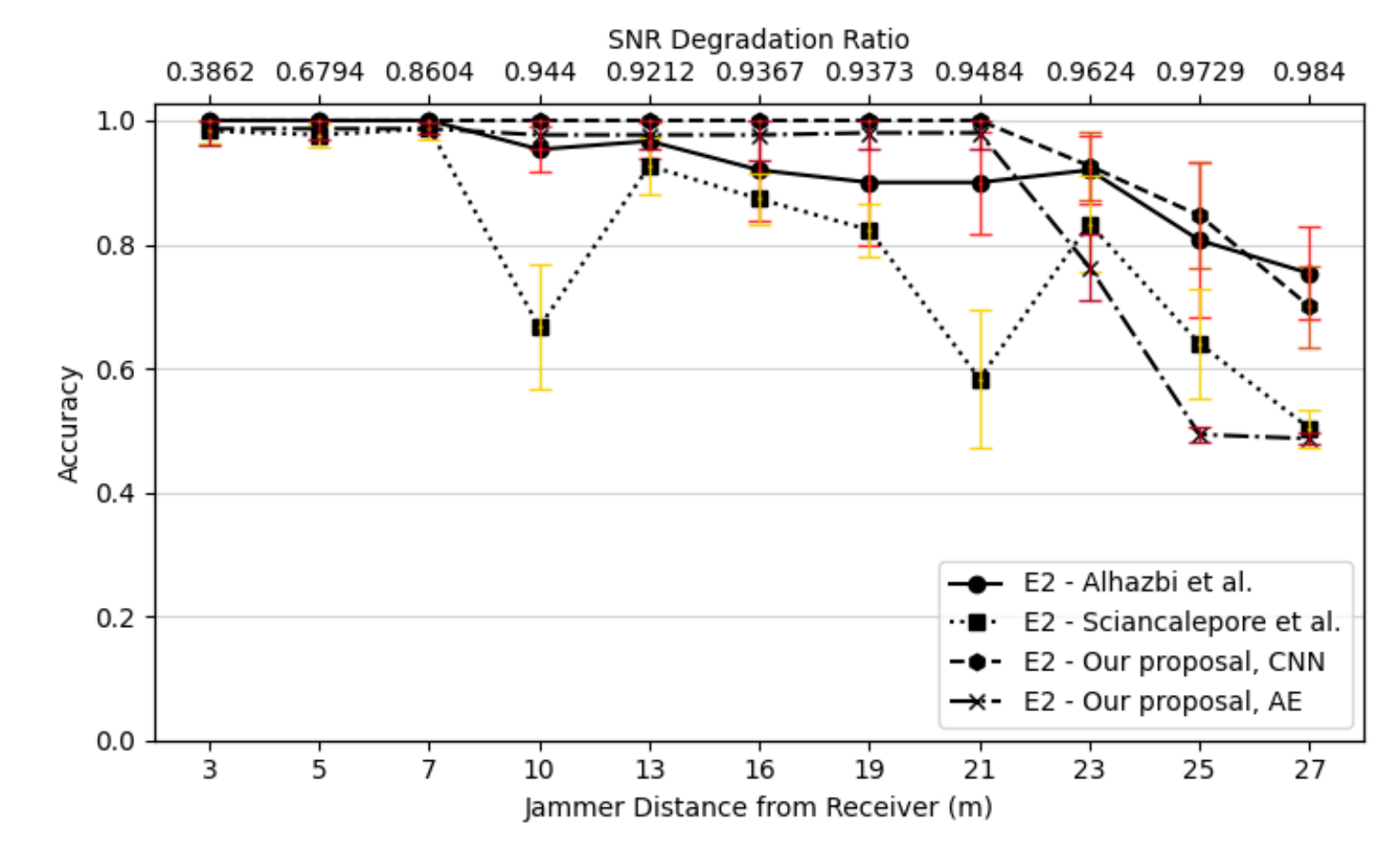}
    \caption{Classification accuracy varying the \emph{Jammer Distance} in environment \emph{E2}.}
    \label{fig:E2_JDist_Res}
\end{figure}

{\bf Exp. 8: Jammer distance.} We evaluate the impact of the distance of the jammer from the communication link on jamming detection accuracy. Figure~\ref{fig:E1_JDist_Res} and~\ref{fig:E2_JDist_Res} show the results for the jamming detection accuracy of the \ac{AE} and \ac{CNN}-based approaches at various distances for scenarios \emph{E1} and \emph{E2}, respectively.
All the proposed solutions have comparable performance in the environment \emph{E1}; indeed, the accuracy is $1$ for all the distances up to $23$~m, where all the detection techniques have a drop in performance. In fact, $23$~m is the distance where the signal power under jamming becomes comparable to the power of the legitimate signal at the receiver, i.e., $SNR_{DR} \approx 1$. In scenario \emph{E2}, we observe a worse performance for the \ac{AE} (E2 - Sciancalepore et al.~\cite{sciancalepore2024_iotj} and E2 - Our proposal, AE).
AE-based detection techniques are less prone to discriminate between interference and jamming, and we note that state-of-the-art solutions are more sensitive to this problem than ours.
In general, the accuracy of our solutions is affected by the distance, starting at $23$ meters in both outdoor environments. In fact, at such distances, the jamming signal is comparable to the legitimate one ($SNR_{DR} = 0.96$), making it difficult to detect.

\subsection{Communication Parameters}
\label{sec:comm_params}

In this subsection, we investigate the performance of jamming detection considering various communication parameters, i.e., the modulation scheme used by the legitimate communication link, the \ac{SNR}, the receiver distance from the jammer, and the mobility patterns of the receiver.

\begin{figure}[h]
    \centering
    \includegraphics[width=0.48\textwidth]{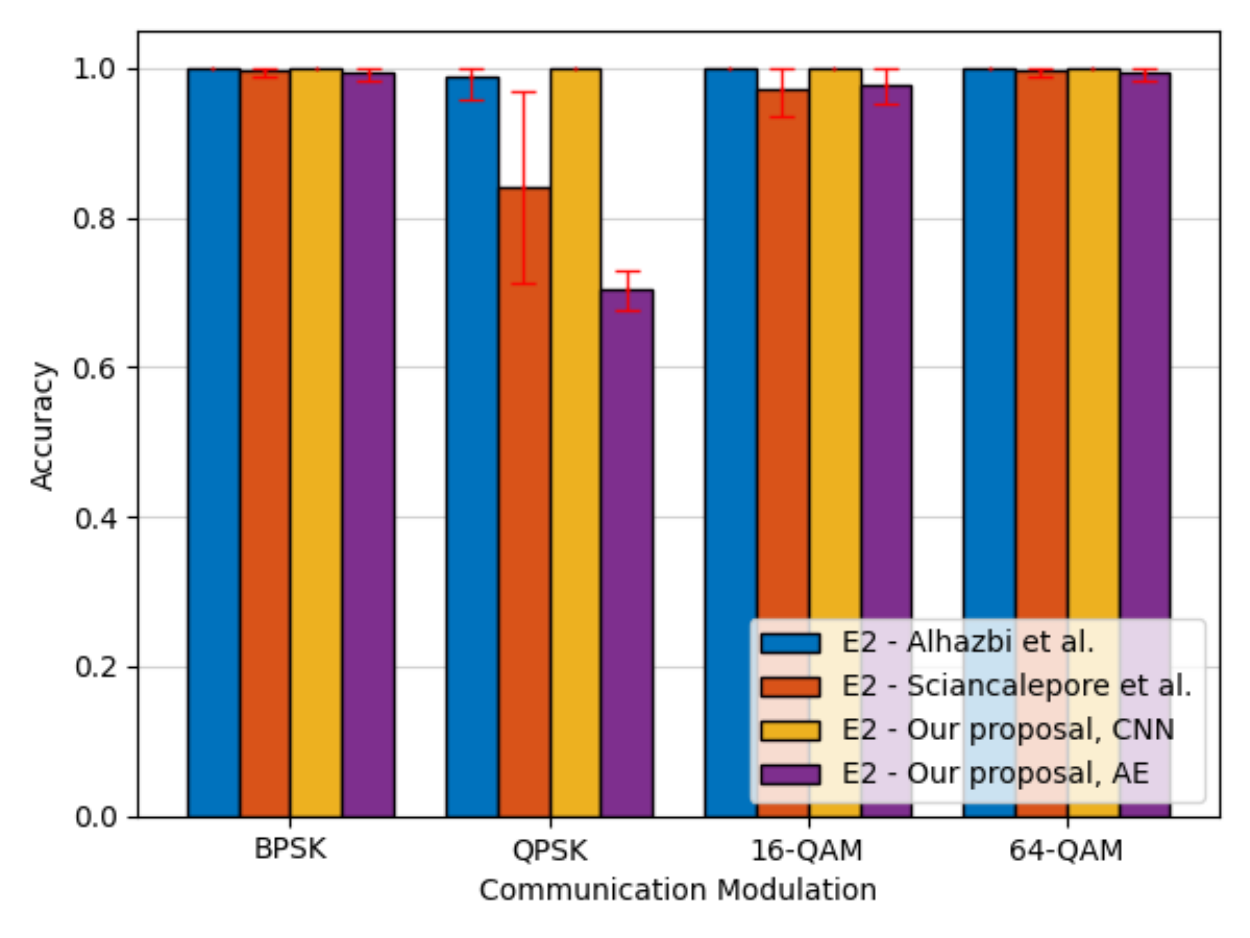}
    \caption{Classification accuracy for the \ac{AE} and \ac{CNN}-based approaches varying the \emph{Communication Modulation Scheme} in environment \emph{E2}.}
    \label{fig:Mod_Res}
\end{figure}

{\bf Exp. 9: Modulation Scheme.} We consider different modulation schemes for the legitimate communication link among the supported options in IEEE 802.11g, i.e., \ac{BPSK}, \ac{QPSK}, 16-\ac{QAM} and 64-\ac{QAM}. We consider the setup shown earlier in Fig.~\ref{fig:Ref_SNR_Mod}, with the TX, RX, and Jammer gain values set to $66$, $50$, and $16$, respectively. Note that we chose the setup and gain values such that the \ac{BER} for the modulation 64-\ac{QAM}---the most complex modulation scheme with the least margin for error---is as low as possible. To adapt the image generation method proposed in the literature to higher-order modulation schemes, we change the x-axis and y-axis limits to center the created clouds while keeping the same rationale. For BPSK, we use $[0, 2]$ and $[-1, 1]$; for QPSK, we use $[-0.293, 1.707]$ and $[-1, 1]$; for 16-QAM, we use $[-0.867, 2.133]$ and $[-1.5, 1.5]$, and finally, for 64-QAM, we use $[-0.883, 2.117]$ and $[-1.5, 1.5]$. Figure~\ref{fig:Mod_Res} shows the results of our experiments, considering both jamming detection approaches and the reference solutions in~\cite{alhazbi2023_ccnc} and~\cite{sciancalepore2024_iotj}, in the environment \emph{E2}.

The accuracy is consistent independently of the modulation scheme and the adopted detection technique. We notice a decrease in the accuracy for QPSK when using the \ac{AE}-based approaches. This drop occurs because the selected (discrimination) threshold is too high, leading to some misclassifications. The \ac{CNN}-based approaches can mitigate such limitations through the apriori knowledge of samples acquired on a jammed communication channel during training, showing their potential for enhanced jamming detection when applicable.

\begin{figure}[h]
\centering
    \subfloat[Scenario E2]{\includegraphics[width=\columnwidth]{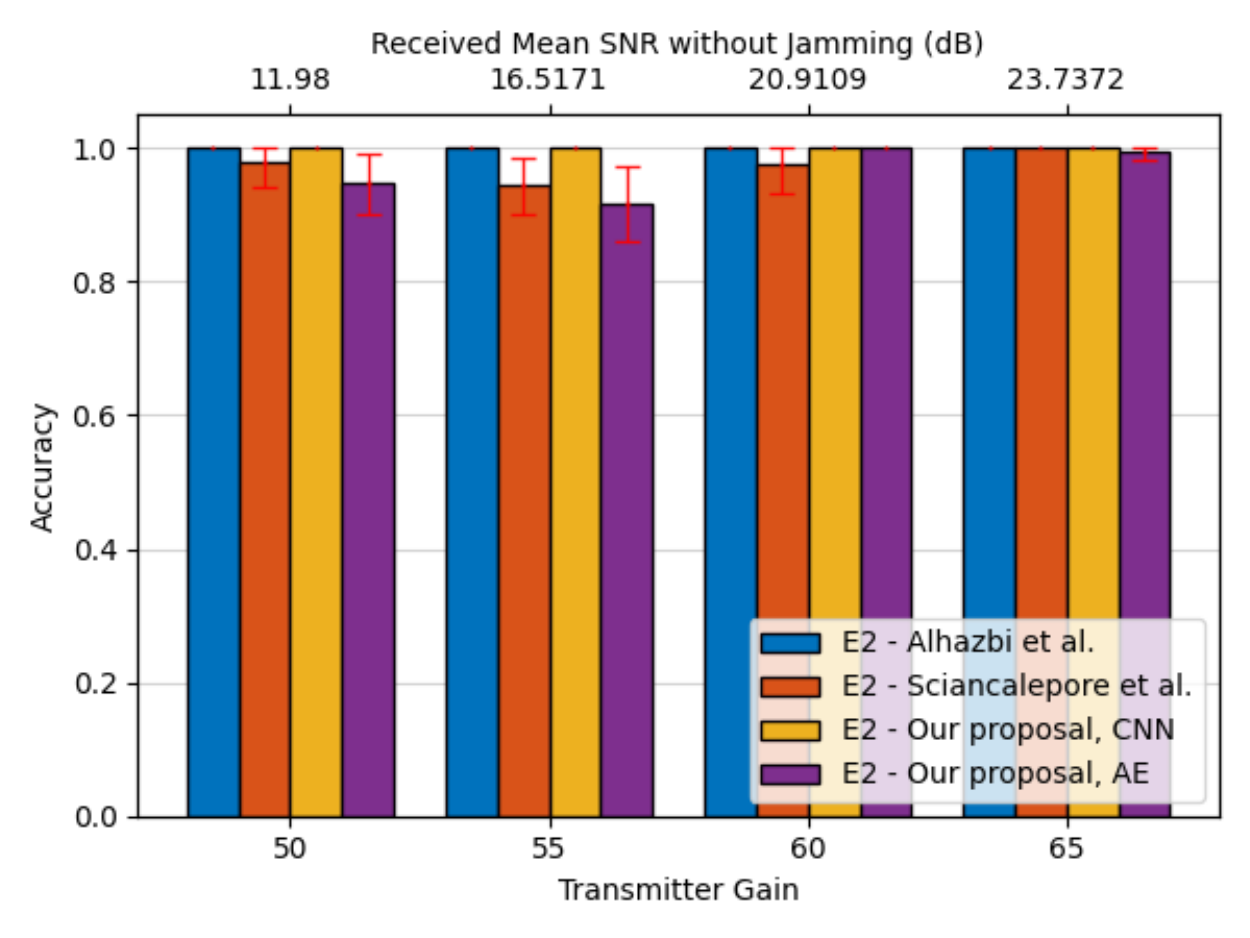}}
    \hfill
    \subfloat[Scenario E3]{\includegraphics[width=\columnwidth]{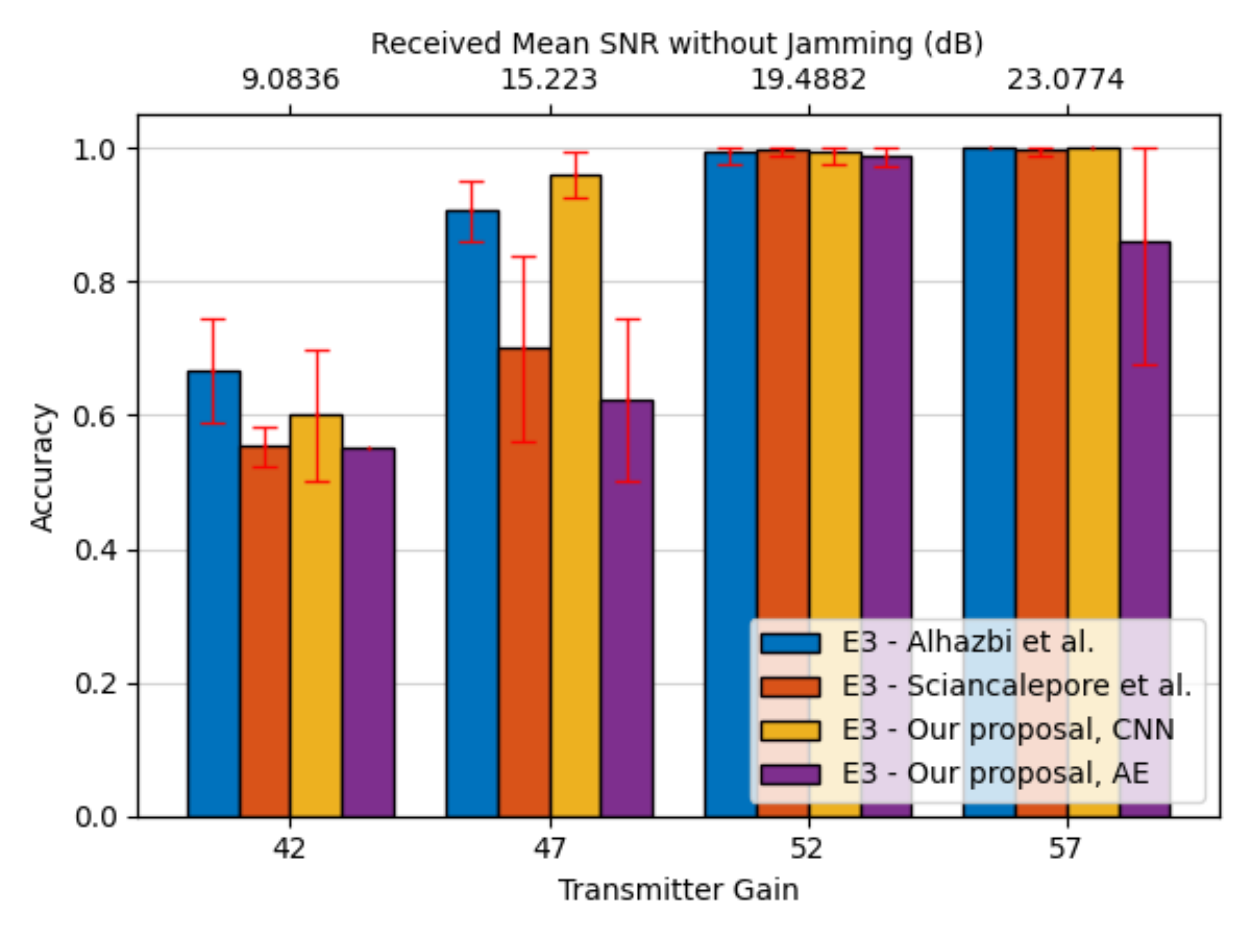}}
    \caption{Classification accuracy varying the \emph{Transmitter Gain}.}
    \label{fig:SNR_res}
\end{figure}

{\bf Exp. 10: Low \acf{SNR} Scenarios.} We also investigate the impact of \ac{SNR} on jamming detection. Since we cannot directly control the \ac{SNR} on a real-world wireless channel, we indirectly influenced it using the gain value set at the transmitter. We use the setup shown earlier in Fig.~\ref{fig:Ref_SNR_Mod}, with gain values of $50$ and $14$ for the receiver and jammer, respectively. Due to the difference in signal propagation, we use different transmitter gains per environment: $[50, 55, 60, 65]$ for E2 and $[44, 49, 54, 59]$ for E3. We selected the lower \ac{SNR} values such that the \ac{BER} under jamming is less than 1\% higher than the \ac{BER} without jamming. We present the results for \emph{E2} in Fig.~\ref{fig:SNR_res}(a) and the results for \emph{E3} in Fig.~\ref{fig:SNR_res}(b).

Low SNR values do not significantly affect detection accuracy in an outdoor environment, as shown in Fig. \ref{fig:SNR_res}(a)---these findings are in line with those of \cite{tedeschi2021_spaccs}. Slow fading, i.e., attenuation of the signal over long distances, also does not affect jamming detection accuracy. Fast fading, i.e., the variability of the noise affecting the \ac{RSS} at a given distance, causes a more significant variance for \ac{RSS} at more considerable distances. Therefore, the detection of jamming is not bound by the low \ac{SNR} of the benign scenario. In contrast, the indoor environment (E3) affects the detection accuracy. As shown in Fig.~\ref{fig:SNR_res}(b), all detection approaches have a low detection accuracy for low SNR values. Thus, we can conclude that the jammer signal affects outdoor communications in a less destructive way due to the reduced multipath fading.

\begin{figure}[h]
    \centering
    \includegraphics[width=\columnwidth]{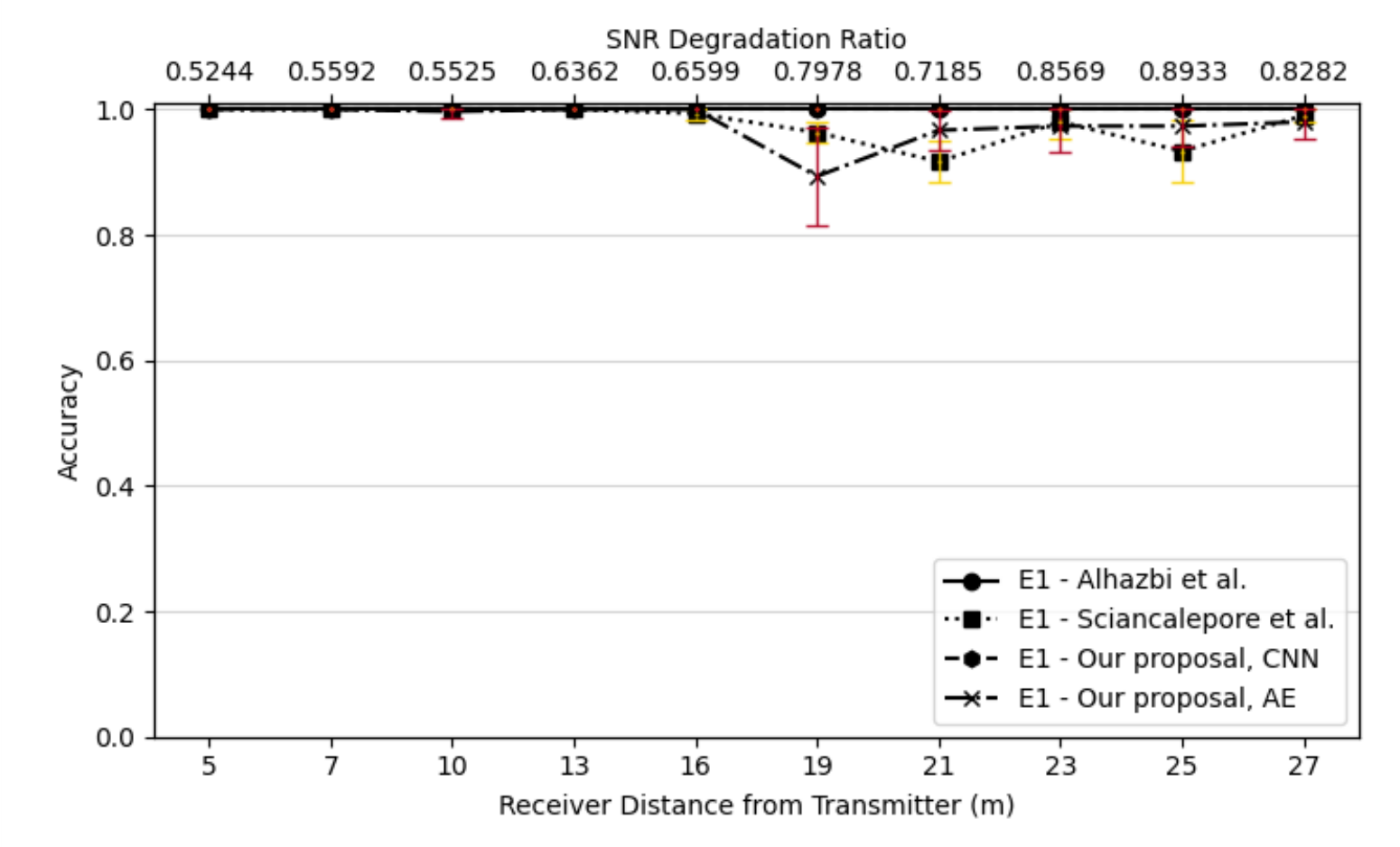}
    \caption{Classification accuracy varying the \emph{receiver distance} in environment \emph{E1}.}
    \label{fig:E1_RDist_Res}
\end{figure}

\begin{figure}[h]
    \centering
    \includegraphics[width=\columnwidth]{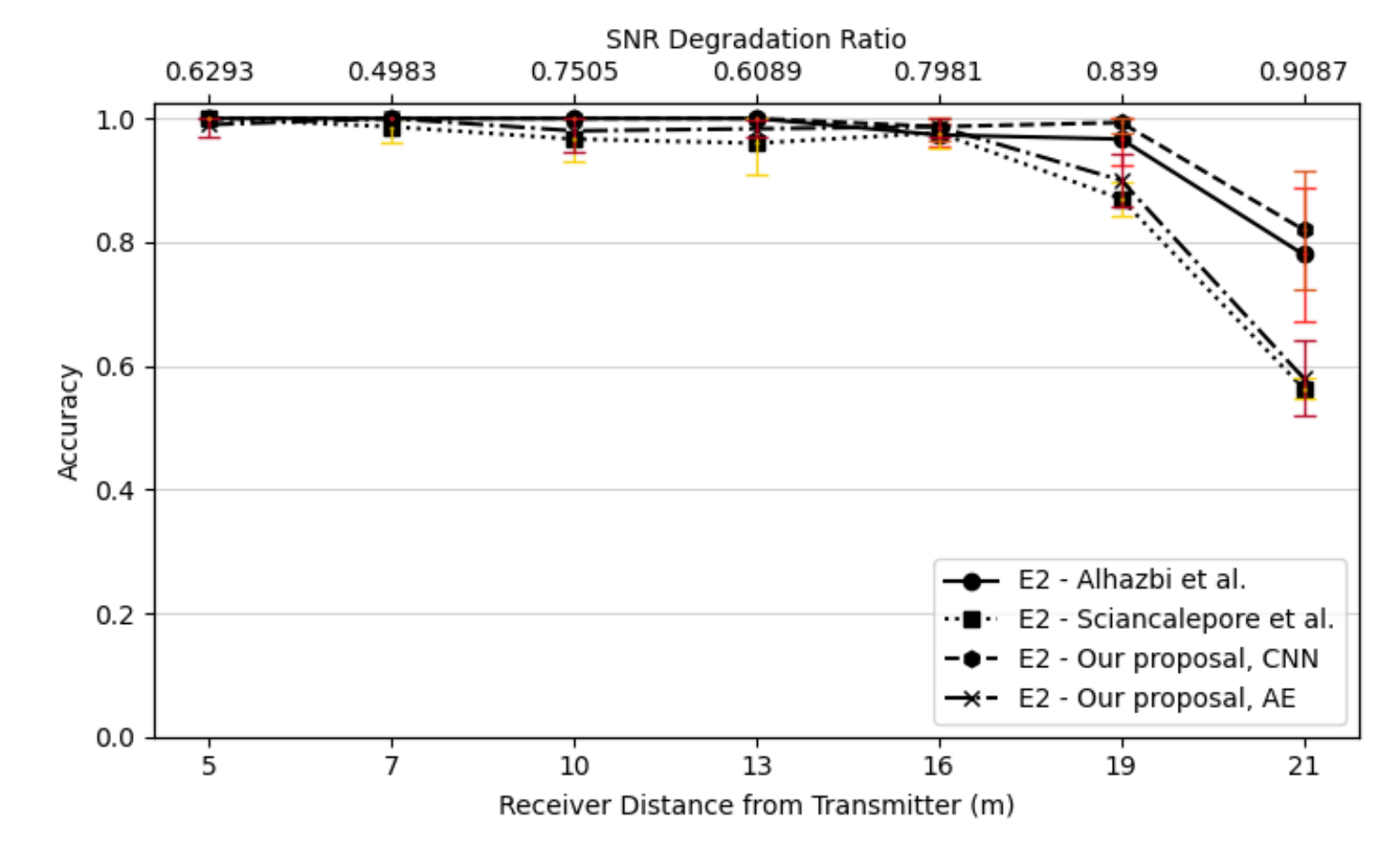}
    \caption{Classification accuracy varying the \emph{receiver distance} in environment \emph{E2}.}
    \label{fig:E2_RDist_Res}
\end{figure}

{\bf Exp.11: Receiver Distance.} To analyze the importance of the SNR of the legitimate communication link for weak-jamming detection, we investigate the impact of the receiver distance from the transmitter on jamming detection accuracy. We present the results in Fig. \ref{fig:E1_RDist_Res} (environment \emph{E1}) and Fig. \ref{fig:E2_RDist_Res} (environment \emph{E2}).
Jamming detection accuracy decreases as the effect of the jammer on the communication channel decreases due to the increased distance between the jammer and receiver. For scenario \emph{E2}, the transmitter and receiver tests with a distance greater than 21 meters were not possible due to the reliability of the communication link at these distances. We experienced significant fluctuations in \ac{SNR}, most likely due to multipath propagation.

\begin{figure}[h]
\centering
    \subfloat[Measurement playground]{\includegraphics[width=.8\columnwidth]{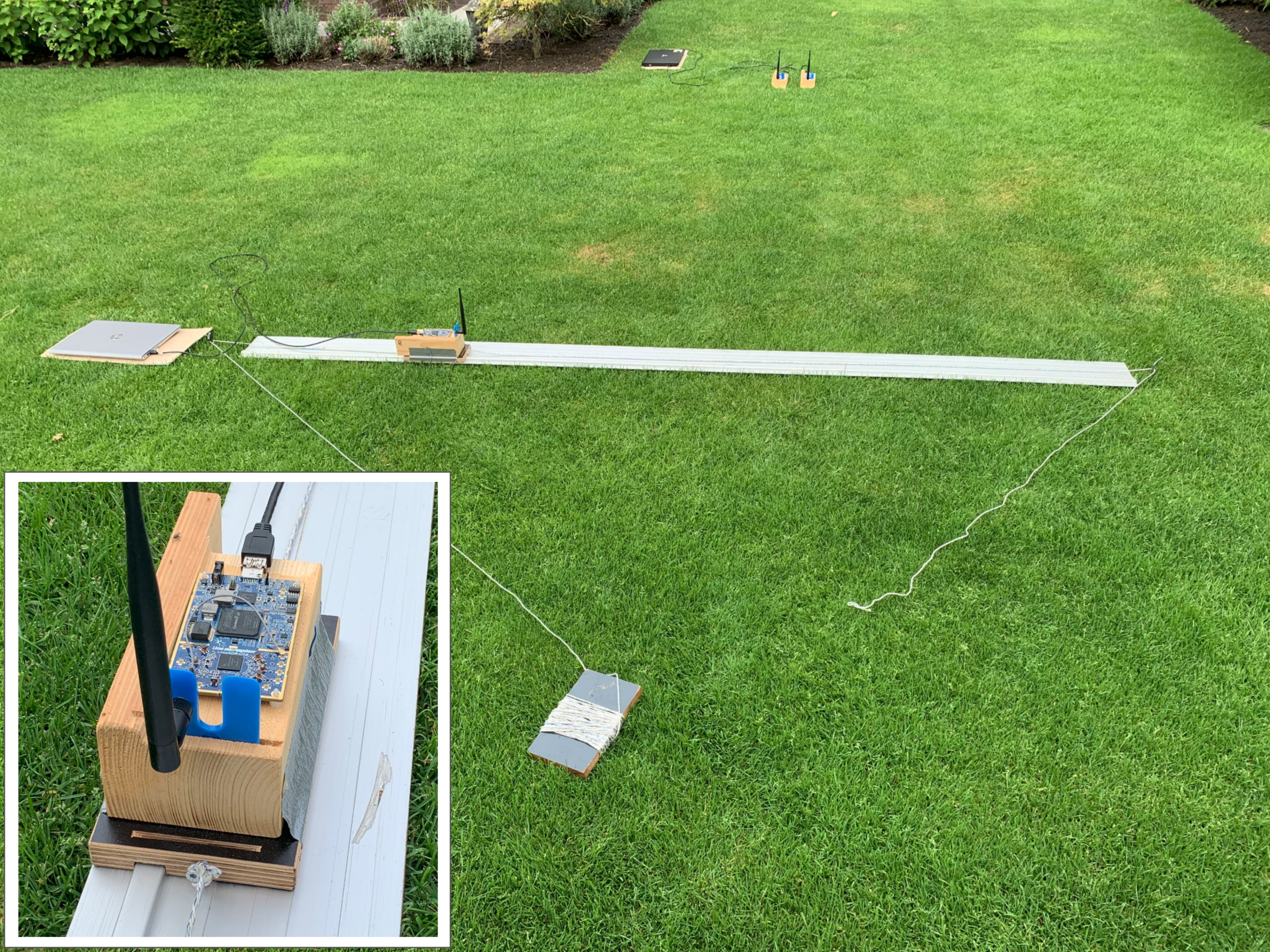}}
    \hfill
    \subfloat[Mobility patterns]{\includegraphics[width=\columnwidth]{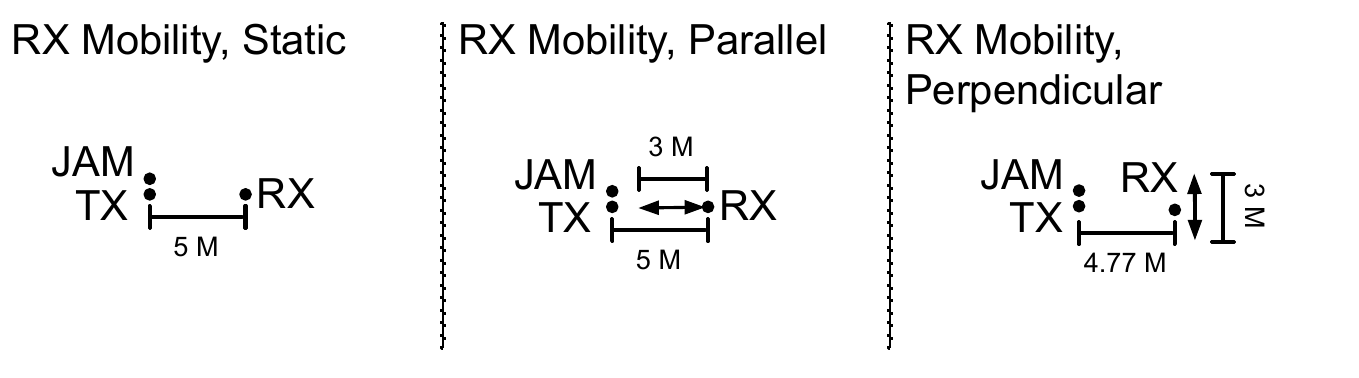}\newline\newline}
    \caption{(a) Setup used for receiver mobility patterns and (b) movement patterns used for the experiments on the impact of receiver mobility on jamming detection. Two tent pegs at both ends of the guide rail allow us to move the receiver across the guide rail using two ropes. The close-up (bottom left) shows the cutout made in the wood to be compatible with the rail.}
    \label{fig:RX_Mobility}
\end{figure}

\begin{figure}[h]
    \centering
    \includegraphics[width=\columnwidth]{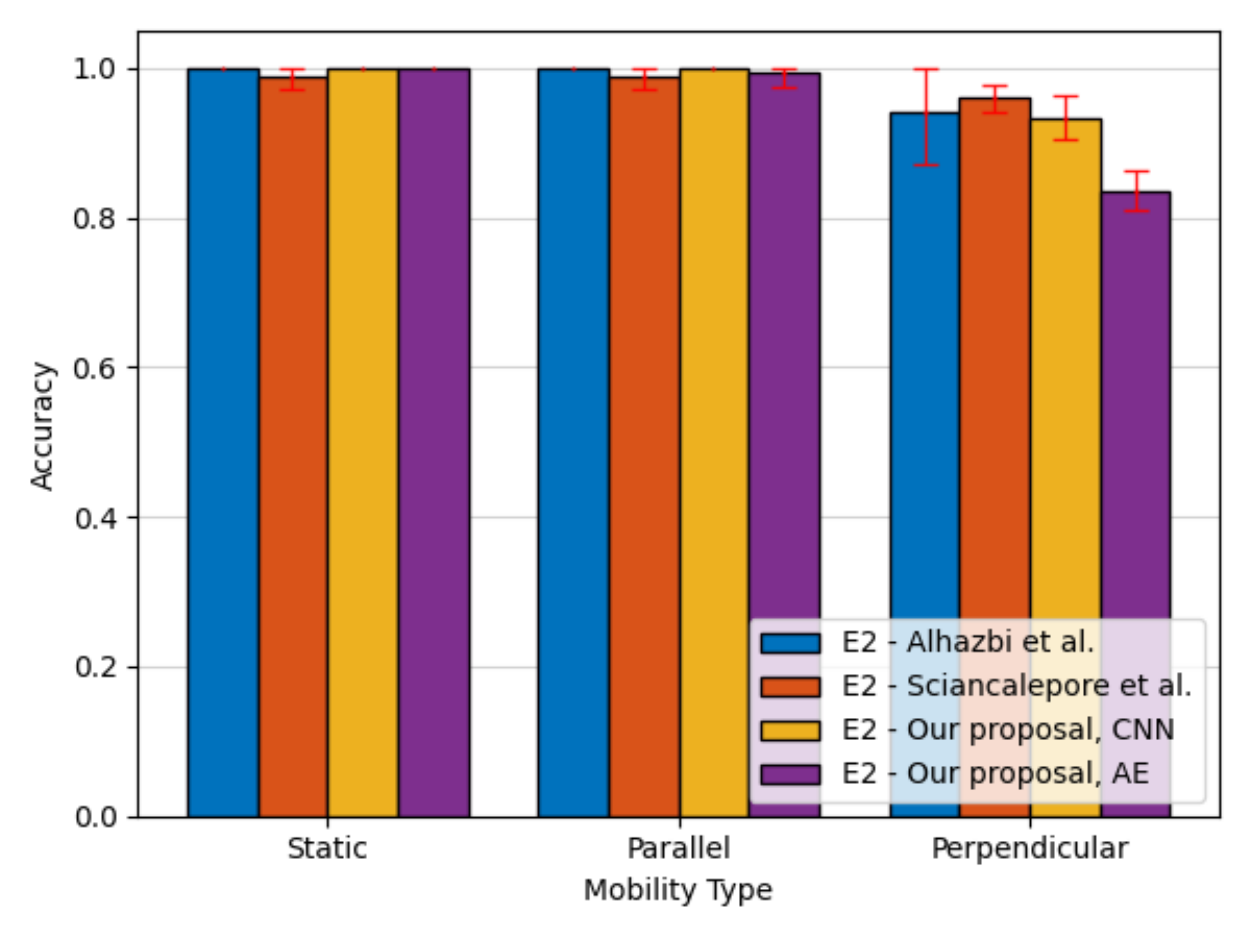}
    \caption{Classification accuracy with various receiver movement patterns in the environment \emph{E2}.}
    \label{fig:Mob_Rx_Res}
\end{figure}

{\bf Exp.12: Receiver mobility.} Finally, we experiment with different mobility patterns, investigating jamming detection accuracy while the receiver moves. For this experiment, we used the gain values $62$, $50$, and $44$ for TX, RX, and jammer, respectively. 
\textcolor{black}{The physical setup used for these experiments is depicted in Fig.~\ref{fig:RX_Mobility}(a) while Fig.~\ref{fig:RX_Mobility}(b) shows the three different mobility patterns considered: static, parallel, and perpendicular, respectively. The user moves up and down a 3-meter-long guide rail with a speed of approximately 12 meters per minute (2 times up and down every minute), bounded by our ability to keep it moving at a constant pace. The receiver is moved by pulling thin ropes attached to both ends of the receiver. }
Figure~\ref{fig:Mob_Rx_Res} shows the results of our tests for the three considered movement patterns.

We observe a slight drop in performance (accuracy) for setup \#3, i.e., the perpendicular movement of the receiver to the transmitter. This $10\%$ drop, primarily relevant for the autoencoder using our solution for image generation, is likely due to the jammer having little effect on the communication link and becoming hardly detectable. Overall, our results confirm that receiver mobility does not affect the accuracy of jamming detection with any of the considered approaches.

\textcolor{black}{\subsection{Overhead evaluation} \label{sec:overhead}
Finally,, we consider the overhead introduced by our solution in terms of delay (processing time) and memory footprint. As previously discussed, our solution has been implemented in MatLab 2023a and run on an AMD Ryzen 9 5900 12-Core Processor, 3.00 GHz, 128GB RAM. Table~\ref{tab:overhead} shows the breakdown of the overhead components in terms of time to generate an image ($\tau_{img}$), time to test an image considering either the CNN-ResNet18 ($\tau_{cnn}$) or the \ac{AE} ($\tau_{ae}$), respectively, the memory footprint for the CNN and \ac{AE} models, i.e., $M_{cnn}$ and $M_{ae}$, respectively, and finally, the memory footprint associated with an image, as a function of the number of samples per image.
\begin{table}[!h]
\footnotesize
\centering
\color{black}
\begin{tabular}{|l|c|c|l|}
\hline
Samples per Image & \multicolumn{1}{c|}{$10^4$} & \multicolumn{1}{c|}{$5\cdot10^4$} & $10^5$ \\ \hline
$\tau_{img}$ (ms) & 10                          & 10                                & 20     \\ \hline
$\tau_{cnn}$ (ms) & 13                          & 55                                & 110    \\ \hline
$\tau_{ae}$ (ms)  & 3                           & 5                                 & 8      \\ \hline
$M_{cnn}$ (MB)    & 45     & 45           & 45     \\ \hline
$M_{ae}$ (MB)     & 14     & 14           & 14     \\ \hline
$M_{image}$ (KB)  & 151KB  & 151KB        & 151KB  \\ \hline
\end{tabular}
\caption{Overhead components of our solution.}
\label{tab:overhead}
\end{table}
Overall, given the reference system considered in this work, the computation overhead sums up to (less than) 130ms and 28ms when considering CNN or Autoencoders, respectively, and the maximum number of samples per image  (worst case), i.e., $10^5$ The memory footprint is independent of the number of samples per image but only depends on the type of neural network considered. We also observe that the image pre-processing technique has a negligible impact on the overall memory footprint since the generated images are very small (compared to the memory required by the model). The major overhead comes from the memory requirement to store the model, i.e., 14MB and 45MB for the Autoencoder and CNN, respectively. Conversely, the memory requirements for the image pre-processing are negligible and sum up to 151KB. Finally, we observe that the number of samples per image affects the processing time, and in particular, CNNs perform much worse than autoencoders (they are more than 4 times slower, also due to running binary classification rather than one-class classification).}

\section{Discussion}
\label{sec:discussion}

{\bf Jamming Detection Approaches.} Our results demonstrate that both \acp{AE} and the chosen \ac{CNN} (Resnet-18) are reliable for weak-jamming detection. The two approaches exhibit similar performance when considering most jamming and communication parameters. However, the \ac{AE}-based technique only requires \emph{not-jammed samples} for the training process, i.e., the samples acquired during regular communications, which are easier to collect in practice. The \ac{CNN}-based approach exhibits better performance when considering some image generation parameters (samples per image and training set size) and some \emph{boundary conditions}, e.g., when the \ac{RSS} of the legitimate communication link becomes comparable to the one of the jammer at the receiver side. However, being a binary classification approach, it requires training on samples from both classes (jammed and not-jammed \ac{IQ} samples), which could be challenging to collect in many real-world scenarios. 
\textcolor{black}{Our experiments considered several measurement parameters potentially impacting jamming detection. Many of such parameters are independent, but some of them are correlated: the jammer distance, transmitter gain, and receiver distance are all correlated through the \ac{SNR} at the receiver. By comparing the results for these experiments among them, it is clear that the \ac{SNR} is the (direct or indirect) variable mostly affecting the capability of detecting weak-jamming: when the SNR degrades significantly or is too low, all investigated jamming detection approaches struggle to detect jamming, while performances improve with lower SNR degradation ratios or higher absolute SNR values.  
}

\textcolor{black}{
{\bf Comparison across Measurement Environments.} The primary aim of this research is to compare various jamming detection methods in specific environments. In general, each environment is characterized by different propagation conditions, which require different fine-tuning of the transmission power and jammer power necessary to achieve weak jamming. In turn, such different configurations make systematic comparison among environments hard to achieve in a fair way across all experiments.
However, as a by-product, our extensive experimental analysis also allows limited comparison of particular methods across scenarios. For example, the results for Experiment 4 (Jammer Oversampling Ratio), Experiment 6 (Jammer Signal Type) and Experiment 7 (Jammer Location) demonstrate that jamming detection across various \ac{JOR} values, jammer hardware and type is always possible independently of environment type. We also notice that multipath makes jamming detection harder, especially in low-SNR conditions. This is evident by comparing Fig.~\ref{fig:E1_JDist_Res} with Fig.~\ref{fig:E2_JDist_Res} and Fig.~\ref{fig:SNR_res} (a) with Fig.~\ref{fig:SNR_res} (b): for both experiments, the presence of multipath leads to more unstable performances, indicating that multipath plays a role in making jamming detection more challenging. Future work could take these results as a starting point to execute a more systematic comparison of performances across measurement environments.
}

{\bf Image generation.} When considering image generation techniques, state-of-the-art solutions from ~\cite{alhazbi2023_ccnc} and~\cite{sciancalepore2024_iotj} exhibit some notable drawbacks. When using \acp{AE}, as proposed in~\cite{sciancalepore2024_iotj},  jamming detection requires more images in the training phase to be reliable ($120$ images vs. $9$ of our proposal, Fig.~\ref{fig:i1_i2_TrainSize_CNN_AE}). In addition, the solution in ~\cite{sciancalepore2024_iotj} is less resilient to interference (see Fig.~\ref{fig:E2_JDist_Res}). Although our solution performs slightly worse for some configurations (e.g., Fig. \ref{fig:Mob_Rx_Res}), overall, it performs more consistently than previous proposals and is compatible by design with any digital modulation scheme.

{\bf Limited Acquisition Bandwidth.} For our experiments, the use of the LimeSDR USB forced us to use an acquisition bandwidth of $5$~MHz instead of $20$~MHz, as required by IEEE 802.11g. This reduced sample rate is a worst-case scenario for jamming detection at the \ac{PHY} layer, since less information is available to the receiver for signal classification. Thus, using larger acquisition bandwidth does not affect our findings.

\textcolor{black}{
{\bf Limited Experiments Area.} We acknowledge that the maximum distance considered for our experiments, i.e., 23~meters, may not be sufficiently representative of extensive jamming in large outdoor scenarios. This limitation of our study is mainly due to two factors. First, the outdoor areas available for our real-world experiments is limited, and we cannot legally perform extensive jamming outdoor without affecting the quality of WiFi services available to other users. Second, the equipment used for the experiments includes \acp{SDR}, which are not professional jammers capable of disrupting communications for hundreds of meters. Such devices have a limited transmission power and, in turn, they can jam nearby communications in a limited space.
On the one hand, such a limitation motivates the need for additional work examining larger jamming distances, which could enhance further the practical relevance and completeness of our study.}
On the other hand, we highlight that the maximum distance considered in our experiments (23~meters) is higher than the only contribution in the current literature investigating the impact of the jammer distance on the detection accuracy, i.e., the contribution in~\cite{sciancalepore2024_iotj}, which considered a maximum distance of $21$~meters, only in an indoor environment.\\
Moreover, although the existence of the limitations mentioned above, we remark that such limitations do not affect the validity of our findings and analysis. In fact, the aim of our analysis is to investigate the effectiveness of detecting weak jamming using various techniques proposed in the literature, and to find the boundary conditions where such techniques could detect weak jamming. As shown in Fig.~\ref{fig:E1_JDist_Res} and Fig.~\ref{fig:E2_JDist_Res}, when the jammer is located 23~meters away from the receiver, all the investigated jamming detection approaches are characterized by sub-optimal performances, indicating that they cannot detect weak jamming reliably. As highlighted through the upper x-axis in those figures, these performances occur with a value of the SNR degradation ratio of 1.0 for the environment E1 (Fig.~\ref{fig:E1_JDist_Res}) and 0.984 for the environment E2 (Fig.~\ref{fig:E2_JDist_Res}). Similar considerations apply for Fig.~\ref{fig:E2_RDist_Res}. Therefore, the usage of more powerful devices over larger distances does not provide additional insight to our analysis and does not affect the validity of our conclusions. 

\textcolor{black}{{\bf Limited Training Data.} For our analysis, we use up to $150$ images of the \emph{NO JAM} dataset for training and an equal number of \emph{JAM} images, depending on the amount of jamming data available for the particular experiment. On the one hand, the usage of such a reduced number of images is an added value of our results. On the other hand, the number of images used for training is relatively limited. Thus, practitioners aiming to deploy our solution(s) in the wild should extend such a training dataset considering domain-specific datasets, collected directly in the environment where the solution is intended to be applied.
}

{\bf Jamming strategies.} In our tests, in line with the relevant literature, we evaluated jamming detection considering \ac{AWGN} and deceptive jamming. Our results are applicable independently from the number of frequencies targeted by the jamming, being it a spot jamming, sweep jamming, barrage jamming, or periodic jamming. However, the adversary could use more advanced types of jamming, e.g., reactive jamming. We first observe that continuous communication between the transmitter and the receiver translates reactive jamming into constant jamming, falling back into the analysis of this paper. When communication is not continuous, in line with the findings in~\cite{oligeri2022_tifs}, our solution is applicable when there are enough \emph{JAM} samples part of an image to enable the correct image classification. The exact duration of the (reactive) jamming signal leading to detection always depends on the sample rate of the jammer and communication link and follows the results provided in Sec.~\ref{sec:results}.

{\bf Known Distance TX-RX.} All the investigated approaches could require knowledge of the distance between the transmitter and receiver. Concerning the mobile scenario introduced in Sec.~\ref{sec:scenario_adv_model}, mobile vehicles usually know such a distance from the communication link with the remote transmitter, e.g., through the adopted telemetry protocols. We recall that our solutions detect jamming early, i.e., when the jamming on the receiver is weak and the communication link is still active. Thus, information availability is never a problem, contrary to \emph{ex-post} approaches in the literature.

\textcolor{black}{
{\bf Scaling to Multi-channel and Multi-hop 802.11 Deployments.} When considering extensive 802.11 deployments encompassing multiple channels and several hops separating IEEE 802.11 nodes, additional challenges apply. To name a few, neighboring 802.11 links could share partially overlapping channels, requiring extensive training to distinguish jamming from unintentional interference, even more with modulation schemes that admit interference by design (e.g., Non-Orthogonal Multiple Access --- NOMA). Moreover, such complex networks require coordination mechanisms to detect and communicate jamming effects at scale. Thus, future research could investigate how to scale up this research to such scenarios.
}

\section{Conclusion and Future Work}
\label{sec:conclusion}
In this paper, we have presented an experimental evaluation of weak-jamming detection solutions, i.e., applying when the power of the jammer at the receiver is low, and the \acl{BER} of the communication link is still not affected by the jammer. To this aim, we have extended our previous solutions to make them work reliably with an actual communication technology used by \acl{COTS} devices, i.e., IEEE 802.11, and with the complex modulation schemes used in such a standard, i.e., OFDM with BPSK, QPSK, 16-QAM, and 64-QAM. We have also enhanced the state-of-the-art methodology for jamming detection through a new image generation approach, generalizing better to complex real-world modulation schemes. Thanks to an extensive experimental campaign using low-cost \acp{SDR}, we demonstrated the effectiveness of our enhanced solution in three heterogeneous wireless environments (indoor, outdoor with multipath, outdoor with minimal multipath) across an extensive range of jamming and communication parameters. 
As part of our future work, we plan to deploy our solution on real \acp{COTS} to evaluate the computational burden and energy overhead it incurs on such constrained devices. \textcolor{black}{We will also investigate to what extent a sudden change in the movement type affect jamming detection capabilities.}

\bibliographystyle{IEEEtran}
\bibliography{main}

\end{document}